\documentclass[journal,draftcls,onecolumn]{IEEEtran}
\pdfoutput=1

\usepackage{graphicx}  
\usepackage{amssymb,amsmath}
\usepackage{tikz}\usetikzlibrary{arrows,shapes,trees}
\usepackage{algorithm}		
\usepackage{algpseudocode}
\usepackage{cite}

\DeclareMathOperator{\Po}{Po}
\DeclareMathOperator{\Bi}{Bi}

\title{Spatially-coupled Split-component Codes with Iterative Algebraic Decoding}

\author{
Lei~M.~Zhang, \IEEEmembership{Student Member,~IEEE} 
Dmitri~Truhachev, \IEEEmembership{Member,~IEEE} and 
Frank~R.~Kschischang, \IEEEmembership{Fellow,~IEEE}
\thanks{L. M. Zhang and F. R. Kschischang are with the Department of ECE, University of Toronto, Toronto, Canada. D. Truhachev is with the Department of ECE, Dalhousie Univeristy, Halifax, Canada. Part of this work was presented at the IEEE Symposium on Information Theory, Hong Kong, June 2015.}
}

\newtheorem{definition}{Definition}
\newtheorem{lemma}{Lemma}
\newtheorem{proposition}{Proposition}
\newtheorem{theorem}{Theorem}

\begin{document}
\maketitle

\begin{abstract}
We analyze a class of high performance, low decoding-data-flow error-correcting codes suitable for high bit-rate optical-fiber communication systems. A spatially-coupled split-component ensemble is defined, generalizing from the most important codes of this class, staircase codes and braided block codes, and preserving a deterministic partitioning of component-code bits over code blocks. Our analysis focuses on low-complexity iterative algebraic decoding, which, for the binary erasure channel, is equivalent to a generalization of the peeling decoder. Using the differential equation method, we derive a vector recursion that tracks the expected residual graph evolution throughout the decoding process. The threshold of the recursion is found using potential function analysis. We generalize the analysis to mixture ensembles consisting of more than one type of component code, which provide increased flexibility of ensemble parameters and can improve performance. The analysis extends to the binary symmetric channel by assuming mis-correction-free component-code decoding. Simple upper-bounds on the number of errors correctable by the ensemble are derived. Finally, we analyze the threshold of spatially-coupled split-component ensembles under beyond bounded-distance component decoding.
\end{abstract}


%
%

\section{Introduction}

\subsection{Staircase Codes and Braided Block Codes}

Staircase codes \cite{smith:2012} are error-correcting codes designed for
communication over the binary symmetric channel (BSC) with thresholds at
high code rates, under hard-decision iterative algebraic decoding, within 0.5 dB of the capacity of an equivalent
binary-input AWGN channel.
Their low decoding data-flow is well-suited to
high bit-rate optical-fiber transmission systems, where communication bit rates of 200
Gb/s or higher have challenged the use of modern message-passing based
capacity-approaching codes, which require orders of magnitude higher decoding data-flow. 

A staircase code consists of many interconnected copies of systematic
binary linear $(n_c,k_c,d_c)$ \emph{component codes}, each of rate
$R_c=k_c/n_c > 1/2$. Its structure consists of a semi-infinite chain of
blocks $B_{i}$, $i=0,1,\dots$, each containing $n_c^2/4$ bits.

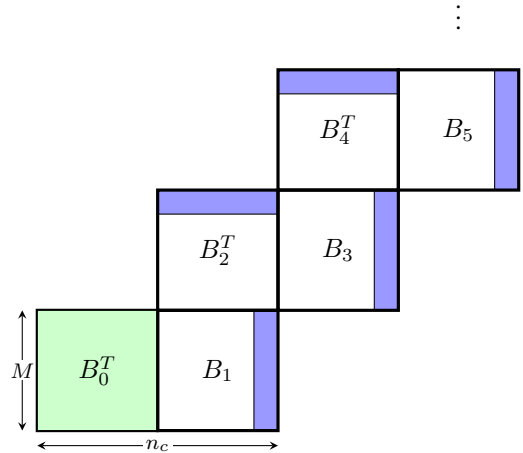
\begin{figure}
\centering
\begin{tikzpicture}[scale=0.8,>=stealth]
		\draw[arrows=<->](-0.25,0)--(-0.25,2);
		\node[fill=white,font=\footnotesize,inner sep=1pt] at (-0.25,1) {$M$};
		\draw[arrows=<->](0,-0.25)--(4,-0.25);
		\node[fill=white,font=\footnotesize,inner sep=1pt] at (2,-0.25) {$n_c$};

		\draw[fill=blue!40] (3.6,0) rectangle (4,2);
		\draw[fill=blue!40] (2,3.6) rectangle (4,4);
		\draw[fill=blue!40] (5.6,2) rectangle (6,4);
		\draw[fill=blue!40] (4,5.6) rectangle (6,6);
		\draw[fill=blue!40] (7.6,4) rectangle (8,6);
		
		\foreach \i in {0,2,4}
			\draw[very thick] (\i,\i) rectangle (\i+2,\i+2);
		\draw [fill=green!20] (0,0) rectangle (2,2);
		\foreach \i in {0,2,4}
			\draw[very thick] (\i+2,\i) rectangle (\i+4,\i+2);
		\foreach \i in {0,2,4}
			\node (\i) at (\i+1,\i+1) {$B_\i^T$};
		\foreach \i in {1,3,5}
			\node (\i) at (\i+2,\i) {$B_\i$};
		\node (6) at (7,7) {$\vdots$};

	\end{tikzpicture}
\caption{Staircase code block structure. Information bits (white) and parity bits (shaded) are shown. Bits in block $B^T_0$ are fixed.}
\label{fig:staircase_blocks}
\end{figure}

Encoding of staircase codes is initialized by filling block $B_0$ with
zeros.
Throughout the encoding process, for $i \geq 1$, the
first $k_s=k_c-n_c/2$ columns of $B_{i}$ are filled with $k_sn_c/2$
information bits. The remaining columns of $B_i$ are filled with
parity-check bits via systematic component code encoding across each row of the matrix
$[B_{i-1}^T \,B_i]$. Hence, the bits of each component code are divided or
\emph{split} across two consecutive blocks. Fig. \ref{fig:staircase_blocks}
illustrates the staircase code block structure and the location of information
and parity bits in each block.

Decoding is performed by an iterative \emph{sliding window decoder}
containing $D$ received blocks, starting from $Y_{0}, \ldots, Y_{D-1}$. For
each index $i$ within the window, the decoder forms the matrix [$Y_i^T\,Y_{i+1}$] and decodes its
rows using any standard syndrome-based, bounded-distance component-code decoder.
As bits are flipped throughout the decoding, syndromes are updated by
bit-masking, costing at most $\lfloor (d_c-1)/2 \rfloor$ operations for every
$n_c$ bits, at every iteration. The process is repeated for a certain number of iterations.
The window then slides by shifting out the decoded $Y_0$ and shifting in
$Y_D$, continuing the process indefinitely.

Braided block codes (BBCs) \cite{truhachev:2003,feltstrom:2009} are a related class of codes
with good performance under iterative algebraic decoding. Each component code 
in a BBC is split into
three or more sections, filled either with information or parity bits.
Encoding follows a similar procedure initiated by two or more blocks of zeros.
The sliding window decoder is used for decoding. A modified braided block
code was proposed in \cite{jian:2013} for high bit-rate optical-fiber
communication systems, with performance and decoding data-flow similar to
staircase codes.

In this paper, we formalize the structure and decoding of staircase codes
and braided block codes by defining the class of \emph{spatially-coupled
split-component codes}.  
We focus on the binary erasure channel (BEC) in Secs.~\ref{sec:ensemble}-\ref{sec:genEnsem} in order to facilitate the analysis.
In Sec.~\ref{sec:ensemble}, we define the spatially-coupled split-component ensemble.
In Sec. \ref{sec:potThresStairEns} we derive a vector recursion which tracks
the ensemble's performance throughout the decoding process and use a
potential function technique \cite{pfister:2014} to determine its
threshold. 
In Sec.~\ref{sec:genEnsem}, we define and analyze ensembles
having a mixture of different component codes.

We focus on the binary symmetric channel (BSC) in Secs.~\ref{sec:bscApproximation}-\ref{sec:genCompDec}. 
In Sec.~\ref{sec:bscApproximation}, we approximate the BSC performance
of spatially-coupled split-component ensembles by applying the BEC analysis under a mis-correction-free decoding assumption. 
In Sec.~\ref{sec:genCompDec}, we analyze spatially-coupled split-component
ensembles over the BSC, exploiting the fact that the component-code decoders can
decode some error patterns beyond their unique decoding radius. 

\subsection{Related Work}

The asymptotic performance of product codes under iterative algebraic decoding was studied in \cite{justesen:2011}. 
The threshold of generalized LDPC (GLDPC) ensembles 
under the BEC with hard-decision message-passing decoding 
 and algebraic component-code decoding was studied in \cite{tang:2007,fossorier:2008}.
The threshold of spatially-coupled generalized LDPC (SC-GLDPC) 
ensembles under the BSC with hard-decision message-passing decoding and algebraic component-code decoding
was studied in \cite{jian:2012}, using the potential function method given in \cite{pfister:2014}. A spatially-coupled LDPC ensemble with structured coupling was studied in \cite{olmos:2015}, with focus on the finite-length performance of the ensemble.

The analyses of hard-decision message-passing decoding in \cite{fossorier:2008,jian:2012}
both focus on message-passing algorithms
where an outgoing message on an edge is independent of the incoming message. 
This \emph{extrinsic} message-passing decoder (EMP) can be analyzed by using density evolution \cite{rutext:2008}.
However, EMP has a more complicated rule governing bit-corrections and requires additional storage of channel-output bit values.

The iterative algebraic hard-decision decoder analyzed in this paper is an \emph{intrinsic}
message-passing decoder (IMP), having the advantages of low decoding data-flow and minimal storage.
However, its analysis is complicated by the dependence between incoming and outgoing messages at each node. The prior analysis of \cite{fossorier:2008,jian:2012} is not directly applicable to the analysis of staircase codes, BBCs, and their generalizations under IMP.

\subsection{Notation}

In this paper, we denote the set of natural numbers including $0$ by $\mathbb{N}$. For any positive integer $a$, we denote the set $\{ 0, 1, \ldots, a-1 \}$ by $[a]$. For any integer $i$, we denote the set $\{i,i+1,\dots,i+a-1\}$ by $i+[a]$. 
We use the summary notation 
\[
\sum_{\sim i_k} v(i_0,\dots,i_{w-1})
\]
to denote
\[
 \sum_{i_0}\dots\sum_{i_{k-1}}\sum_{i_{k+1}}\dots\sum_{i_{w-1}} v(i_0,\dots,i_{w-1})
\]
the sum over index vectors $(i_0,\dots,i_{w-1})$ of length $w$ with the $k$th component fixed.

The indicator function $\mathbb{I}[P]$ indicates the truth of predicate $P$, taking
value $1$ when $P$ is true, and taking value $0$ when $P$ is false.

We denote a binomial probability mass function with parameters $(n,p)$ for $n\geq 0$, $p\in[0,1]$
as 
\[
\Bi[n,p](i) = \binom{n}{i} p^i (1-p)^{n-i},\, i=0,1,\dots,n.
\] 

Similarly, we denote a Poisson probability mass function with mean $\alpha>0$
as 
\[
\Po[\alpha](i) = e^{-\alpha}\frac{\alpha^i}{i!},\, i=0,1,\dots.
\]

%
%

\section{Ensemble Definition}\label{sec:ensemble}

\subsection{Spatially-coupled Split-component Codes}

In this section, we define an ensemble of graphs that
represent spatially-coupled split-component codes. Throughout this
section, all component codes are fixed to some
binary linear $(n_c,k_c,d_c)$ code $\mathcal{C}$.
The \emph{coupling width} of the ensemble, is denoted
by $w$, where $w \geq 2$, and we assume
that $w$ is a divisor of $n_c$.
The \emph{spatial length} of the ensemble, denoted
by $L$, is the total number of blocks $B_i$ that we consider
in an ensemble. The meaning of these parameters will become clear in the ensemble definition.

Define $k\in [L]$ to be an index of spatial location, or simply
\emph{index}.  Associated with each index are
$M$ \emph{constraint nodes} and $N$ \emph{variable nodes}.
We associate $n_c$ \emph{half-edges} with each constraint node 
and $v$ half-edges with each variable node. 
In other words, each constraint node is of degree $n_c$ and 
each variable node is of degree $v$.
The parameters $M$, $N$, $v$ and $w$ will need to
satisfy certain relations, given below.

The half-edges incident on each constraint node at index $k$
are partitioned or \emph{split} deterministically into $w$ edge-types,
so that there are exactly $n_c/w$ half-edges of each type $\tau \in [w]$.
Since a constraint node represents a component code of length $n_c$,
a split constraint node represents the partition of a component codeword
$(x_0,x_1,\dots,x_{n_c-1}) \in \mathcal{C}$ according to
\[
(\underbrace{ x_0,\dots,x_{n_c/w-1} }_{\textrm{type }0},
\underbrace{ x_{n_c/w},\dots,x_{2n_c/w-1} }_{\textrm{type }1}, \dots, 
\underbrace{x_{(w-1)n_c/w},\dots,x_{n_c-1}}_{\textrm{type }w-1}).
\]
Hence, a bit-ordering of the component code
must be given in order to specify an ensemble. 
We also assume an arbitrary but fixed labelling of the constraint nodes at each spatial-location by the set $[M]$. Every constraint node half-edge in the ensemble is specified by three parameters: spatial index, constraint node label, and bit label.

The collection of all $Mn_c /w $ half-edges of type $\tau$ at index $k$
is referred to as the constraint edge-bundle, or simply \emph{bundle}, of type $\tau$ at index $k$. Each half-edge in this bundle will be connected to some variable half-edge at index $k-\tau$.

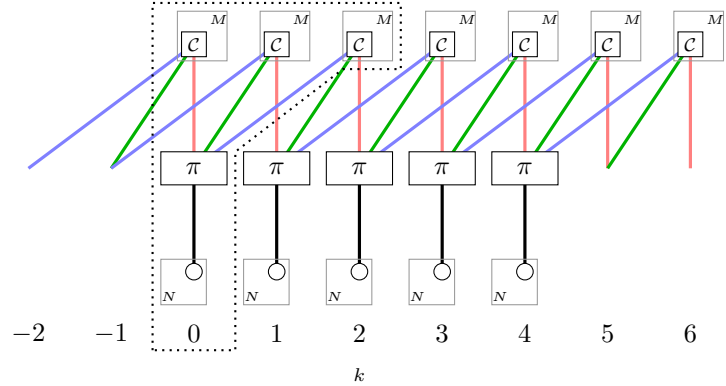
\begin{figure}
\centering
%
%
%

\begin{tikzpicture}[scale=0.11,>=stealth]
	\foreach \i in {-2,-1,0,1,...,6}			
		\node (\i) at (10*\i,5) {$\i$};
	\node at (20,0) {\scriptsize$k$};

	\foreach \i in {0,10,...,60}		
	{
		\draw [very thick,color=red!50] 	(\i,40) -- (\i+0,25);			
		\draw [very thick,color=black!30!green] (\i,40) -- (\i-10,25);			
		\draw [very thick,color=blue!50] 	(\i,40) -- (\i-20,25);			
	}

	\foreach \i in {0,10,...,40}		
	{
		\draw [very thick] (\i,12.5) -- (\i,25);
		\draw[fill=white] (\i-4,25-2) rectangle (\i+4,25+2);
		\node (\i) at (\i,25) {$\pi$};
		
		\draw[color=black,fill=white] (\i,12.5) circle (1);
		\draw[color=gray!80] (\i+1.5,12.5+1.5) rectangle (\i-4,12.5-4);
		\node at (\i-3,12.5-3) {\tiny $N$};
	}

	\foreach \i in {0,10,...,60}		
	{
		\draw[color=black,fill=white] (\i-1.5,40-1.5) rectangle (\i+1.5,40+1.5);
		\node (\i) at (\i,40) {\scriptsize$\mathcal{C}$};
		\draw[color=gray!80] (\i-2,40-2) rectangle (\i+4,40+4);
		\node at (\i+3,40+3) {\tiny $M$};
	}

	\draw[thick,dotted] (-5,3) -- (-5,45) -- (25,45) -- (25,37) -- (18,37) -- (5,27) -- (5,3) -- (-5,3);

\end{tikzpicture}




		

	
\caption{Illustration of a $\mathcal{E}(\mathcal{C},2,M,3)$ spatially-coupled split-component ensemble. Plate notation is used to indicate a group of multiple variable/constraint nodes at a spatial location. Each edge represents an edge-bundle containing $Mn_c/3$ and $2N$ edges incident on constraint and variable nodes respectively. Un-connected edges represent suppressed edge-bundles. Elements of the ensemble completely enclosed by the dotted line represent the single-interleaver ensemble at spatial-location $0$ as defined in Sec. \ref{subsec:single_interleaver_ens}.}
\label{fig:ensemble_graph}
\end{figure}

At each index $k$, there are a total of $Nv$ half-edges incident on variable nodes to be connected to $w$ constraint edge-bundles originating from indices $k+\tau$ for $\tau\in[w]$. The balance between constraint and variable half-edges gives rise to the constraint $N = Mn_c/v$, assuming $M$ is chosen so that $N$ is an integer. At each spatial location, we make connections between the variable and constraint half-edges by using an interleaver of size $Nv$ (equivalently $Mn_c$). Fig. \ref{fig:ensemble_graph} illustrates the corresponding graph for the case when $w=3$.

A variable half-edge inherits the type of the constraint half-edge it is connected to via the interleaver. One interleaver is associated with the group of variable nodes at each spatial location. We define the variable edge-bundle of type $\tau$ at an index $k$ to be the collection of all half-edges incident on variable nodes at index $k$ which inherited type $\tau$. Since $w$ constraint edge-bundles of distinct types entered each interleaver, there are exactly $w$ variable edge-bundles of distinct types $\tau\in[w]$ at each index $k$.

Because we have assumed a finite spatial length $L$, the connection
pattern described above must be modified near the ends of the graph.
We introduce $w-1$ \emph{termination} indices $k \in \{ L, ..., L+w-2\}$
with which we associate only constraint nodes.  
When connecting a constraint-bundle of type $\tau$ at any position $k$,
if $k-\tau<0$ or $k-\tau\geq L$, the corresponding variable half-edge
bundle is absent.  Therefore, these constraint bundles are simply
deleted, and the corresponding constraint is modified by
assuming that the missing bits have value zero.  We refer to
such deleted edge-bundles as \emph{suppressed edge-bundles}.
For termination, bits in spatial locations $k \in \{ L-w+1, \ldots, L-1 \}$
must be chosen for compatibility with the modified constraints.

The design rate of the ensemble is given by
\[ 
R(L) = 1-\frac{LM(n_c-k_c)+(w-1)M(n_c-k_c)}{LN}.
\]
It can be obtained by considering a parity-check matrix $H$
which corresponds to an arbitrary sample graph in the ensemble. The maximum number of linearly independent rows in $H$ is $(L+w-1)M(n_c-k_c)$, which is an upper-bound on its rank. The rate is given by the dimension of the null-space of $H$, normalized by the number of variable nodes $LN$. 

Note that in the case of parallel edge-connections between a variable node and a constraint node, e.g. $m$ parallel edges, the column vector in $H$ corresponding to the variable node contains the mod-$2$ sum of $m$ columns of the component code parity-check matrix, selected according to the interleaver connections. Hence the presence of parallel edges does not affect the design rate. 

As $L \rightarrow \infty$ the design rate converges to
\begin{equation}\label{eqn:rate}
R =  1-v(1-R_c). 
\end{equation}
Throughout this paper, we assume that $L$ is some fixed large value. Therefore, we only consider the rate given by (\ref{eqn:rate}) and will not indicate dependence on $L$ in our notation. Furthermore, we only consider positive design rates, which requires the additional constraint on the rate of the component code
\[
R_c > 1 - \frac{1}{v}.
\]
Note that the asymptotic design rate of a spatially-coupled split-component ensemble is the same as the design rate of the uncoupled GLDPC ensemble with the same component code and variable degree \cite{fossorier:2008}\cite[Theorem 1]{tanner:1981}.

For a fixed $(n_c,k_c,d_c)$ component code $\mathcal{C}$,
by allowing all possible permutations at the edge-bundle
interleavers, the ensemble
$\mathcal{E}(\mathcal{C},v,M,w)$ of graphs describing
spatially-coupled split-component codes is obtained.
A $\mathcal{E}(\mathcal{C},2,M,3)$ ensemble is shown in Fig. \ref{fig:ensemble_graph}.

A staircase code with component code $(n_c,k_c,d_c)$ (assuming $n_c$ is even) is an element of the spatially-coupled split-component ensemble with parameters $M=n_c/2$, $w=2$, $v=2$, $N=(n_c/2)^2$. Each interleaver connects one half-edge of a variable node at index $k$ to a half-edge in the constraint edge-bundle of type $0$ at index $k+0$. It connects the other half-edge of a variable node to a half-edge in the constraint edge-bundle of type $1$ at index $k+1$. It is easy to verify that braided block codes are also elements of spatially-coupled split-component ensembles with the appropriate parameters and interleavers.

\subsection{Decoupling interleavers}

Within our ensemble definition, there exist interleaver realizations which may result in a spatially \emph{uncoupled} chain of component codes. Consider an interleaver in which the half-edges incident on the same variable node are all connected to the same constraint half-edge bundle, i.e. all half-edges incident on a variable node have the same edge-type. Such an interleaver is necessary for an uncoupled sample graph. Moreover, a sample graph containing $w-1$ such interleavers at consecutive spatial-locations is uncoupled. Here we consider the probability of sampling one such interleaver uniformly at random from the set of all possible interleavers and show that such an event becomes exceedingly rare as $M\rightarrow\infty$. Since interleavers at different spatial-locations are sampled independently, this is an upper-bound on the probability of sampling decoupled sample-graphs from an ensemble.

We refer to an interleaver of size $Mn_c$ (or equivalently $Nv$) as a \emph{decoupling interleaver} if it connects all half-edges incident on the same variable node to constraint half-edges belonging to the same bundle. 

\begin{theorem}\label{thm:decoupInt}
For a $\mathcal{E}(\mathcal{C},v,M,w)$ ensemble with $v\geq 2$, the probability of sampling a decoupling interleaver uniformly from the set of all interleavers of size $Mn_c$ scales as $O(N^{-(v-1)N})$ for $N\rightarrow\infty$.
\end{theorem}
\begin{IEEEproof}
See Appendix \ref{append:decoupInt}
\end{IEEEproof}

\subsection{Related ensembles}

A spatially-coupled split-component ensemble can be viewed as the spatial-coupling of a \emph{multi-edge-type} ensemble \cite[pp. 382-384]{rutext:2008} with generalized check nodes. Note that it is not a protograph ensemble since edge-bundles of different edge-types are mixed within a single interleaver. 

The difference between SC-GLDPC ensembles, such as those defined in \cite{jian:2012}, and spatially-coupled split-component ensembles is the distribution over constraint node degree-types (see definition in Sec. \ref{sec:potThresStairEns}). In SC-GLDPC ensembles, the random assignment of constraint-node edges to adjacent spatial locations results in a multinomial distribution over degree types. Spatially-coupled split-component ensembles, on the other hand, only contain the degree type $(n_c/w,\dots,n_c/w)$.

Olmos and Urbanke in \cite{olmos:2015} define a spatially-coupled LDPC ensemble which shares some similarities with spatially-coupled split-component ensembles. The key difference between these ensemble definitions is that the deterministic splitting of edges occurs at constraint nodes for spatially-coupled split-component ensembles, whereas in \cite{olmos:2015} the deterministic splitting of edges occurs at variable nodes. A consequence of this difference is that our analysis is able to track the average error-probability at each spatial-location. In comparison, \cite{olmos:2015} tracks a global parameter over the entire spatially-coupled chain in order to determine the threshold. Furthermore, the focus of this work is on the analysis of spatially-coupled split-component ensembles under hard-decision algebraic component decoding, instead of the more complex extrinsic message-passing decoder studied in \cite{olmos:2015}.

%
%

\section{BEC Decoding Analysis}\label{sec:potThresStairEns}

\subsection{Preliminaries}

We assume that the all-zero codeword is sent over a
BEC with erasure probability $p$. Since the channel and component decoders are symmetric, this assumption does not affect the decoder performance. A code graph is sampled from a spatially-coupled split-component ensemble by sampling, independently and uniformly at random, one interleaver at each spatial-location. The channel realizations are sampled and edges in the code graph incident on \emph{correctly} received variable nodes are removed, resulting in a \emph{residual graph}. 

Each node $y$ in the residual graph is assigned a vector of length $w$ called the \emph{degree-type} (or simply \emph{degree})
$\deg(y) = (i_0(y),\dots,i_{w-1}(y))$, where the $\tau$th component
$i_{\tau}(y)$ is the number of edges of type $\tau$ incident on $y$.
The \emph{total degree} of $y$ is $\sum_{\tau \in [w]} i_{\tau}(y)$.

We refer to a constraint node as being \emph{recoverable}
if it has total degree in the set
$\{ 1, \ldots, d_c-1 \}$ and as \emph{un-recoverable} if it is not recoverable. 
We refer to an edge as being recoverable (un-recoverable)
if it is incident on a recoverable (un-recoverable) constraint node.
We refer to a variable node as being \emph{recovered} if 
its total degree is smaller than $v$.

Decoding is modelled by Algorithm \ref{alg:peelDecWithVar}, which is a generalization of the iterative algebraic decoding of staircase codes. In this algorithm, $l \in \mathbb{N}$ counts iterations.
The algorithm is also a generalization of the peeling decoder used in the analysis of LDPC ensembles over the BEC \cite{luby:2001},\cite[Sec. 3.19]{rutext:2008}. Hence, we also refer to Algorithm \ref{alg:peelDecWithVar} as a \emph{generalized peeling decoder}.

\begin{algorithm}[t]
\begin{algorithmic}

\State $l\gets 0$

\While{$\exists$ constraint node $c$ such that $c$ is recoverable} 
	\ForAll {recoverable $c$}
         	\State{Remove all edges incident on $c$}
	\EndFor
	\ForAll {recovered variable nodes $x$}
		\State{Remove all edges incident on $x$}
	\EndFor
	\State $l\gets l+1$
\EndWhile

\end{algorithmic}
\caption{Generalized peeling decoder}
\label{alg:peelDecWithVar}
\end{algorithm}

Denote a degree-type by $\underline{i}\triangleq (i_0,\dots,i_{w-1})$ and the set of all degree-types by $D = \{ \underline{i} | 0 \leq \sum_{\tau \in [w]} i_{\tau} \leq n_c \}$. Define the \emph{degree-type distribution} (DD)
\[
\mathcal{R}^k_{(\underline{i})}(l) : [L+w-1] \times \mathbb{N} \times D \mapsto \mathbb{N}
\]
to be a function which tracks the number of constraint nodes at index $k$, at the beginning of decoding iteration $l$, for each degree-type $\underline{i}$. The notation is chosen to reflect the importance of the iteration $l$, relative to the other variables, in the following derivation. 

Initially, when $l=0$, each \emph{component} of the degree-type of a constraint node, averaged over all sample graphs and channel realizations, has a binomial distribution with a mean of $n_c p / w$ (or zero, if the edge-type corresponds to a suppressed edge-bundle at the particular spatial-location).
The distribution of each component is independent of the other components. Hence, the average initial DDs are
\[
\mathbb{E}(\mathcal{R}^{k}_{(\underline{i})}(0)) = M \prod_{\tau \in [w]} \Bi[n_c/w,p\cdot\mathbb{I}[0\leq k-\tau<L]](i_{\tau}).
\]

\subsection{The Differential Equation Method}

The decoding analysis relies on a technique known as the differential equation method (DEM). It was generalized to the technique used here and applied to a number of problems in graph theory by Wormald \cite{wormald:1995,pittel:1996,wormald:1997}. It was introduced into coding theory for the analysis of irregular LDPC ensembles over the BEC in \cite{luby:1997,luby:2001}. Subsequently, it has been successfully used in the analysis of finite-length scaling behaviour over the BEC for LDPC ensembles \cite{amraoui:2009} and spatially-coupled LDPC ensembles \cite{olmos:2015}, and in the analysis of a number of message-passing algorithms \cite{olmos:2011,yi:2008}. Historically, the DEM was replaced by the less complex density evolution method \cite{rutext:2008} for the analysis of many message-passing algorithms over graphs. However, it remains important in the cases where density evolution does not apply. Our current problem is one such case.

Given a sequence of discrete-time stochastic processes indexed by $m\in \mathbb{N}$ and a corresponding sequence of (finite or countable) state spaces of increasing cardinality, the DEM first derives a set of difference equations describing the expected change, from time step $i$ to $i+1$, in a set of functions defined over the underlying process. For the method to apply, the maximum number of time steps that each process can take must scale with $m$, i.e. $i \in [Cm]$ for a positive constant $C$. Substitute the variables $t\triangleq i/(Cm)$ and $\Delta t \triangleq 1/(Cm)$. As $m\rightarrow \infty$, the difference equations converge to a set of continuous-time differential equations in the variable $t$. Solutions of the differential equations are the limiting expected-evolution of the functions defined over the underlying process. Moreover, Theorem \ref{thm:wormald} (see below) ensures the concentration of the functions of any sample path of the process to the differential equation solutions as $m\rightarrow\infty$.

In the following definition of Theorem \ref{thm:wormald}, we use a number of symbols which have already appeared. We introduce these symbols as variables ``local'' to the theorem which will not be used after the end of the description. Let $\Omega$ be a countable probability space and let $S$ be some finite or countable set. Let $Q = (Q_0,Q_1,\dots)$ be a discrete-time random process where to every $\omega \in \Omega$ corresponds a realization $(Q_0(\omega),Q_1(\omega),\dots)$ of the process with $Q_i(\omega)\in S$. The \emph{history} of the process up to time $t$ is the sequence $H_t = (Q_0,Q_1,\dots,Q_t)$. Define $S^+ \triangleq \cup_{i\geq 1}S^i$ where $S^i = \underbrace{S\times\dots\times S}_{i}$. For a function $y: S^+\mapsto \mathbb{R}$, the random variable $y(H_t)$ is denoted by $Y_t$. We say that a function $f:\mathbb{R}^j \rightarrow \mathbb{R}$ satisfies a \emph{Lipschitz condition} on some sub-set $O\subseteq\mathbb{R}^j$ if there exists a constant $L>0$ such that
\[
|f(u)-f(v)| \leq L \cdot \sum_{i=1}^j |u_i - v_i|
\]
for all $u,v\in O$. 

Consider a sequence of discrete-time random processes indexed by $m$
\[
Q^{(m)} = (Q^{(m)}_0,Q^{(m)}_1,\dots)
\]
with $Q^{(m)}_i \in S^{(m)}$. Let
\[
H^{(m)}_t \triangleq (Q^{(m)}_0,\dots,Q^{(m)}_t)
\]
be the history up to time $t$ of the $m$th random process. The set of histories for the $m$th random process is $S^{(m)+}$. The random variable defined over the history up to time $t$ of the $m$th random process  is $y(H_t^{(m)}) \triangleq Y_t^{(m)}$.

Let $d$ be a positive integer. Let $D\subseteq\mathbb{R}^{d+1}$ be some bounded connected open set containing the closure of 
\[
\{ (t,z_1,\dots,z_d) : \textrm{Pr} ( Y^{(i,m)}_0=z_im, 1\leq i \leq d ) > 0 \textrm{ for some } m \}.
\]
For $1\leq i \leq d$ and all positive integers $m$, let $y^{(i,m)} : S^{(m)+} \rightarrow \mathbb{R}$ be a function such that $|y^{(i,m)}(h) < Cm|$ for all histories $h \in S^{(m)+}$ and for some constant $C$ (independent of $i$, $m$, $h$). Furthermore, let $f_1,\dots,f_d$ be functions from $\mathbb{R}^{d+1}$ to $\mathbb{R}$. Given $D$, we define the \emph{stopping time} $T_{D}(Y_t^{(1,m)},\dots,Y_t^{(d,m)})$ to be the minimum $t$ such that $( t/m,Y_t^{(1,m)},\dots,Y_t^{(d,m)} ) \notin D$.

\begin{theorem}[Wormald's theorem \cite{wormald:1995}]\label{thm:wormald} Suppose that the following conditions are fulfilled:
\begin{enumerate}
\item (Boundedness) There exists a constant $C'$ such that for all $m$, $0 \leq t < T_D$, and $1\leq i \leq d$
\[
| Y^{(i,m)}_{t+1} - Y^{(i,m)}_t | < C';
\]
\item (Trend functions) For all $1\leq i \leq d$ and uniformly over all $(m,t)$ with $0 \leq t < T_D$ we have
\begin{equation*}
\mathbb{E} ( Y^{(i,m)}_{t+1} - Y^{(i,m)}_t | H^{(m)}_t ) \\
= f_i ( t/m,Y^{(1,m)}_t/m,\dots,Y_t^{(d,m)}/m );
\end{equation*}
\item (Lipschitz continuity) For each $1\leq i \leq d$, the function $f_i$ is continuous and satisfies a Lipschitz condition on the intersection of $D$ with the half-space $\{(t,z_1,\dots,z_d) : t \geq 0\}$.
\end{enumerate}

Then the following statements hold
\begin{enumerate}
\item For $(0,\zeta_1,\dots,\zeta_d) \in D$, the system of differential equations
\[
\frac{dz_i}{d\tau} = f_i(\tau,z_1,\dots,z_d),\; i=1,\dots,d
\]
has a unique solution in $D$ for $z_i:\mathbb{R}\rightarrow\mathbb{R}$ passing through $z_i(0)=\zeta_i$, $1\leq i \leq d$. This solution extends arbitrarily close to the boundary of $D$.
\item There exists a strictly positive constant $c$ such that 
\[
\textrm{Pr} (Y^{(i,m)}_t > mz_i(t/m) + cm^{5/6}) < dm^{2/3}\exp(-\sqrt[3]{m}/2)
\]
for $0 < t < \sigma m$ and for each $i$, where $z_i(t)$ is the solution in 1) with $\zeta_i=\mathbb{E}(Y^{(i)}_0)/m$, and $\sigma = \sigma(m)$ is the supremum of those $\tau$ to which the solution can be extended.
\end{enumerate}
\end{theorem}

For a proof of Theorem \ref{thm:wormald} please refer to \cite{wormald:1995,wormald:1997,luby:2001}. In our application, the random process is the evolution of the residual graphs starting from the initial graph. The total number of edges in the initial graph corresponds to the variable $m$. Each time-step in the random process corresponds to one iteration of the generalized peeling decoder, hence variable $t$ corresponds to iteration number $l$. The set $S^{(m)}$ is the set of all possible residual graphs and $S^{(m)+}$ is the set of all possible sequences of residual graphs. For some $m$ and $t$, given $h_t\in S^{(m)+}$ we can enumerate $\mathcal{R}^k_{(\underline{i})}(t)$. The DD is a set of random variables corresponding to $Y_t^{(i,m)}$. The correspondence is established by indexing the set $[L+w-1]\times D$ serially using $0 \leq i \leq |[L+w-1]\times D| \triangleq d$ and considering each $\mathcal{R}^k_{(\underline{i})}(t)$ for fixed $k$ and $\underline{i}$ to be a distinct random variable. 

Derivations of the trend functions and differential equations are given in Sec. \ref{subsec:single_interleaver_ens}. Here we provide a high-level explanation to highlight the connections with Theorem \ref{thm:wormald}. First, we find the expected change in each DD random variable $\mathcal{R}^k_{(\underline{i})}(l)$ from decoding iteration $l$ to $l+1$. The trend functions are obtained by normalizing the expected change in $\mathcal{R}^k_{(\underline{i})}(l)$ by the initial number of edges in the graph, $m$. This results in a set of difference equations. Substitute the variables $u\triangleq l/m$ and $\Delta u \triangleq 1/m$. As $M\rightarrow\infty$ (hence $m\rightarrow\infty$), we obtain a set of differential equations of the form shown in statement 1) of Theorem \ref{thm:wormald}, in the continuous-time variable $u$. Similar derivations in the application of DEM to coding theory can be found in \cite{luby:2001,olmos:2011}, and in the tutorial \cite[Sec. C.4]{rutext:2008}.

We now give an overview of the steps in the decoding analysis. First, we focus on a collection of sub-graphs containing only those edges connected by an interleaver at index $k$ and the variable and constraint nodes on which they are incident (an example is shown by the elements enclosed by the dotted line in Fig. \ref{fig:ensemble_graph}). The analysis of generalized peeling decoding for this collection of sub-graphs is very similar to the peeling decoding analysis of an LDPC ensemble \cite{luby:2001}. To adapt the analysis technique to generalized peeling decoding, we define an \emph{incremental decoder} which removes \emph{one} randomly chosen recoverable edge from all recoverable edges at each iteration. The expected evolution of DDs, averaged over the collection of sub-graphs at index $k$, is obtained by a straightforward extension of the derivation in \cite{luby:2001}.

Next, we consider the parallel generalized peeling decoding of \emph{every} collection of sub-graphs at indices $k\in[L]$. By assuming that the evolution of DDs involved in the sub-graphs at index $k$ are \emph{not} influenced by the decoding at adjacent indices $\{k-w+1,\dots,k-1,k+1,\dots,k+w-1\}$, we effectively decouple the spatially-coupled chain into $L$ distinct collections of sub-graphs, referred to as sub-ensembles. The DD evolution for each sub-ensemble is then given by the previously obtained results at a single index $k$. Decoding at each sub-ensemble must terminate at some \emph{stopping time} dependent on $k$. Note that due to the decoupling assumption, the stopping times at each $k$ are determined by \emph{local} convergence conditions only. We refer to the progression from an initial time to the maximum stopping time over $k\in[L]$ as a \emph{round} of decoding. 

We show that the actual DDs of \emph{un-recoverable} constraint nodes at the end of each round of decoding can be calculated from the DDs of un-recoverable constraint nodes obtained under the decoupling assumption. Furthermore, the updated DDs of un-recoverable constraint nodes determine a \emph{global} convergence condition for the entire spatially-coupled chain. By parametrizing the DDs of un-recoverable constraint nodes, we obtain a vector recursion which describes the expected DD evolution over the spatially-coupled split-component ensemble, under generalized peeling decoding. By using Theorem \ref{thm:wormald}, we conclude that the DD evolution of any sample graph almost surely follows the expected DD evolution as $M\rightarrow\infty$.

\subsection{Single-interleaver ensemble}\label{subsec:single_interleaver_ens}

We focus on the sub-graphs of a spatially-coupled split-component ensemble induced by the edges belonging to an interleaver at an arbitrary spatial-location index $k\in[L]$. This is the collection of sub-graphs containing all variable nodes at index $k$, their edges, all constraint nodes at indices $k+\tau$ for $\tau \in [w]$ and their type $\tau$ edge-bundles (but no edge-bundles of any other type). An example of this collection of sub-graphs is shown in Fig. \ref{fig:ensemble_graph} for $k=0$ as the elements completely enclosed by the dotted line. We refer to this collection as the \emph{single-interleaver ensemble} at index $k$, or simply the \emph{sub-ensemble} at index $k$.

For residual graphs of the sub-ensemble at index $k$, we analyze an incremental version of Algorithm \ref{alg:peelDecWithVar}, which we call an \emph{incremental decoder}. For this decoder, at the beginning of decoding iteration $l$, \emph{one} recoverable edge is randomly selected from the set of all recoverable edges. The chosen edge is removed, creating a recovered variable node of degree $v-1$. This recovered variable node and its edges are also removed to end decoding iteration $l$. The decoder iterates until no recoverable edges remain.

All of the edges and variable nodes which were removed by the incremental decoder would have been removed by Algorithm \ref{alg:peelDecWithVar}, hence the performance of Algorithm \ref{alg:peelDecWithVar} is at least as good as the incremental decoder. Moreover, the incremental decoder terminates if and only if the residual graph is empty or is non-empty and have no recoverable constraint nodes. These are the exact terminating conditions of Algorithm \ref{alg:peelDecWithVar}. Therefore, the performances of these two decoders are identical.

Let $G$ be a sample graph from the sub-ensemble at index $k$. Let $E$ be the total number of edges in $G$, i.e. $E=Nv=Mn_c$, which corresponds to the variable $m$ in Theorem \ref{thm:wormald}. It is also an upper-bound on the number of decoding iterations $l$. Let $Q_l$ be the edge removed by the incremental decoder at iteration $l$. For the moment, we suppress the superscript $(m)$ to simplify the notation.

Since incremental decoding is applied to all sub-ensembles in parallel, each DD at index $k$ is influenced by the decoding of sub-ensembles at other indices. For example, $\mathcal{R}^{k}_{(\underline{i})}(l)$ can be modified by the decoding of sub-ensembles at indices $k-\tau$ for $\tau\in[w]$ (unless the edge-bundle of type $\tau$ was suppressed). We \emph{decouple} the DDs in the spatially-coupled chain by the following assumption. 

Consider the sub-ensemble at index $k$. Each DD at index $k+\tau$, $\tau\in[w]$ contributes a type $\tau$ edge-bundle to the sub-ensemble. We make the assumption that throughout the incremental decoding of the spatially-coupled chain, the DDs are constant  \emph{from the perspective of each sub-ensemble}, except for those edge-bundles involved in the sub-ensemble. In DD notation this condition is given by
\begin{equation}\label{eqn:frozen}
\sum_{i_{\tau}}\mathcal{R}^{k+\tau}_{(\underline{i})}(l) = \sum_{i_{\tau}}\mathcal{R}^{k+\tau}_{(\underline{i})}(0)\end{equation}
for all $\tau\in[w]$ and $l>0$, where the sum is over degree-types with all except the $\tau$th element fixed to  $\underline{i}$. 

As a result, at the start of decoding each $\mathcal{R}^k_{(\underline{i})}(0)$ is copied $w$ times, one for each sub-ensemble at $k-\tau$, $\tau\in[w]$. In order to satsify (\ref{eqn:frozen}), each of these copies must be modified \emph{differently} by the incremental decoding at index $k-\tau$. The resulting DDs are re-combined when we consider the entire coupled chain in Sec. \ref{subsec:coupled_ens}.

Given sample graph $G$ from the sub-ensemble at index $k$ and history $h_l=(q_0,\dots,q_l)$, we define the functions $y_{\tau}^{(i)}$ for $i\in[(n_c/w)^w]$ and $\tau\in[w]$, to be mappings from the residual peeled graph at the beginning of decoding iteration $l$ to $\mathcal{R}^{k+\tau}_{(\underline{i})}(l)$. Here $i$ is an index over the set of all degree types $\underline{i}$ in some arbitrary but fixed order. 

The functions are upper-bounded by $M$. In addition, since at most $v$ degree types of constraint nodes can be modified during one iteration, their differences are strictly bounded by $C'=v+1$. Hence, condition 1) in Theorem \ref{thm:wormald} is satisfied.

A point of confusion may arise due to our two-variable parametrization $y_{\tau}^{(i)}$. There are $w(n_c/w)^w$ such functions, hence $d=w(n_c/w)^w$. Theorem \ref{thm:wormald} uses a single parameter $1\leq i \leq d$ to index these functions. However, since it is more natural to organize these functions according to their membership in DDs at different indices $k+\tau$, we shall continue to use the two-variable parametrization.

Since DDs are a Markov process under peeling decoding \cite{luby:2001}, the history $H_t$ in the definition of condition 2) can be replaced by DDs at iteration $l$, which we represent using $\Theta_l \triangleq \{ \mathcal{R}^{k+\gamma}_{(\underline{i})}(l)\;, \gamma \in [w]\}$. Let $T = \{ \underline{i} | \sum_{\tau} i_{\tau} \leq d_c-1 \}$ and $S = D \setminus T$ denote the set of recoverable and un-recoverable degree types.

Given $\Theta_l$, let $E_{\tau}$ denote the number of edges of type $\tau$. Let $E^T_{\tau}$ denote the number of \emph{recoverable} edges of type $\tau$. Let $E$ be the total number of edges in the residual graph. On the variable side, let $F$ denote the number of edges and let $F^T$ denote the number of \emph{recoverable} edges. By edge balance we have $F=\sum_{\tau\in[w]} E_{\tau}$ and $F^T=\sum_{\tau\in[w]} E^T_{\tau}$.

The incremental decoder can continue as long as $F^T>0$. If the edges were not separated into distinct edge-types, then one iteration of incremental decoding is equivalent to one iteration of peeling decoding. With distinct edge-types, we must average over all edge-types. Let $\underline{1}_{\tau}$ denote a vector of length $w$ in which all entries are $0$ except at position $\tau$, where it is $1$.

\begin{lemma}\label{lemma:trend}
For all un-recoverable degree types $\underline{i} \in S$
\begin{multline}\label{eqn:trend}
\mathbb{E}(\mathcal{R}^{k+\tau}_{(\underline{i})}(l+1) - \mathcal{R}^{k+\tau}_{(\underline{i})}(l) |\Theta_l) \\
= [- i_{\tau}\mathcal{R}^{k+\tau}_{(\underline{i})}(l)+ (i_{\tau}+1)\mathcal{R}^{k+\tau}_{(\underline{i} + \underline{1}_{\tau})}(l)] \frac{v-1}{F}
\end{multline}
\end{lemma}
\begin{IEEEproof}
See Appendix \ref{subsec:trend}
\end{IEEEproof}

Let $u=l/E$ and define
\[
r^{k+\tau}_{(\underline{i})}(u) = \frac{1}{E}\mathcal{R}^{k+\tau}_{(\underline{i})}(uE)
\]
\[
r^{k+\tau}_{(\underline{i})}(u+1/E) = \frac{1}{E}\mathcal{R}^{k+\tau}_{(\underline{i})}(uE+1)
\]
\[
e(u) =F/E
\]
\[
e^T(u)=F^T/E.
\]
We re-write (\ref{eqn:trend}) as 
\begin{align}
\mathbb{E}(\mathcal{R}^{k+\tau}_{(\underline{i})}(l+1) - \mathcal{R}^{k+\tau}_{(\underline{i})}(l) |\Theta_l) 
&= [- i_{\tau}r^{k+\tau}_{(\underline{i})}(u+1/E)+ (i_{\tau}+1)r^{k+\tau}_{(\underline{i} + \underline{1}_{\tau})}(u)] \frac{v-1}{e(u)}\label{eqn:trend_norm} \\
& \triangleq f_{\tau}^{(i)}(u,r_{(\underline{i})}^{k+\tau}).\nonumber 
\end{align}
\noindent
In a slight abuse of notation, we denote by the last line a function $f_{\tau}^{(i)}$ which contains $d+1$ parameters: $u$ and $r_{(\underline{i})}^{k+\tau}$ for each $\underline{i} \in D$.

To verify condition 3) of Theorem \ref{thm:wormald}, observe that for any $\eta>0$ each expression in (\ref{eqn:trend_norm}) satisfies a Lipschitz condition as long as $e(u)>\eta$. Hence, Theorem \ref{thm:wormald} applies to all steps during the decoding process for which an arbitrarily small fraction of edges remain in the graph.

By Theorem \ref{thm:wormald}, for $\underline{i}\in S$ we have the differential equations 
\begin{multline}\label{eqn:diff_eqns}
\frac{dr^{k+\tau}_{(\underline{i})}(u)}{du} \\
= [- i_{\tau}r^{k+\tau}_{(\underline{i})}(u)+ (i_{\tau}+1)r^{k+\tau}_{(\underline{i} + \underline{1}_{\tau})}(u)] \frac{v-1}{e(u)}
\end{multline} 
with initial conditions
\begin{multline}\label{eqn:init_conds}
r^{k+\tau}_{(\underline{i})}(0) = \frac{1}{n_c}\\
\times \prod_{\gamma\in[w]}\Bi\left[n_c/w,p \cdot \mathbb{I}[0 \leq k + \tau - \gamma < L]\right](i_{\gamma}).
\end{multline}

These differential equations can be solved following the steps in \cite[Appendix B]{luby:2001}. Making a change of variables $x(u)=\exp( -\int_0^u 1/e(s) ds)$, the solutions under the initial conditions (\ref{eqn:init_conds}) for $\underline{i}\in S$ are
\begin{multline}\label{eqn:solutions}
r^{k+\tau}_{(\underline{i})}(x) = \frac{1}{n_c} \Bi\left[\frac{n_c}{w},px^{v-1}\right](i_{\tau}) \\
\times \prod_{\gamma\in [w] \setminus \tau} \Bi\left[n_c/w,p \cdot \mathbb{I}[0 \leq k + \tau - \gamma < L]\right](i_{\gamma}).
\end{multline}
We represent the constant factor (product over $\gamma$) by $K_{\tau}$ for convienence. Note that under the change of variables, the ``time'' variable $u$, which increases with decoding progress, is changed to the variable $x$, which \emph{decreases} with decoding progress.

Finally, by Theorem \ref{thm:wormald} we have
\begin{equation*}
\mathcal{R}^{k+\tau}_{(\underline{i})}(x) = Mr^{k+\tau}_{(\underline{i})}(x) + O(M^{5/6})
\end{equation*}
with probability at least 
\[
1-w(n_c/w)^w (Mn_c)^{2/3}\exp(-\sqrt[3]{Mn_c}/2).
\]
For $M\rightarrow\infty$, the evolution of the (normalized) DDs of residual graphs under incremental decoding is almost-surely given by $r^{k+\tau}_{(\underline{i})}(x)$.

We apply the change of variables to the edge-count functions: $e(x)=e(x(u))$ and $e^T(x)=e^T(x(u))$. The incremental decoder can proceed as long as $e^T(x)>0$. By definition
\begin{align*}
e(x) &= \sum_{\tau\in[w]}\sum_{\underline{i}\in D} i_{\tau} r^{k+\tau}_{(\underline{i})}(x) \\
&= \sum_{\tau\in[w]} \left( \sum_{\underline{i}\in S} i_{\tau} r^{k+\tau}_{(\underline{i})}(x) + \sum_{\underline{i}\in T}  i_{\tau} r^{k+\tau}_{(\underline{i})}(x) \right)\\
&= \sum_{\tau\in[w]} \sum_{\underline{i}\in S} i_{\tau} r^{k+\tau}_{(\underline{i})}(x) + e^T(x)
\end{align*}
hence
\begin{equation}\label{eqn:eTx_from_ex}
e^T(x) = e(x) - \sum_{\tau\in[w]} \sum_{\underline{i}\in S} i_{\tau} r^{k+\tau}_{(\underline{i})}(x).
\end{equation}
For all $x$ such that $e^T(x)>0$, at every decoding iteration $v$ edges are removed from the sub-graph at index $k$. The differential equation for $e(u)$ is simply
\[
\frac{de(u)}{du} = -v.
\]
Its solution in the variable $x$ is
\begin{equation}\label{eqn:ex}
e(x) = e(1)x^v
\end{equation}
with initial condition
\begin{equation}\label{eqn:e_init_cond}
e(1) = \sum_{\tau\in[w]}\sum_{\underline{i}\in D}i_{\tau}r^{k+\tau}_{(\underline{i})}(u=0).
\end{equation}

Let $\hat{S} = S \cup \{\underline{i} | \sum_{\tau}i_{\tau} = d_c-1 \}$. Combining (\ref{eqn:solutions}) -- (\ref{eqn:e_init_cond}) we obtain
\begin{align}
e^T(x) &= e(1)x^v - \sum_{\tau\in[w]}\sum_{\underline{i}\in S}i_{\tau} \frac{1}{n_c} \Bi\left[\frac{n_c}{w},px^{v-1}\right](i_{\tau})  K_{\tau} \nonumber \\
&= \sum_{\tau\in[w]}\left( x^v \sum_{\underline{i}\in D}i_{\tau}\frac{1}{n_c} \Bi\left[\frac{n_c}{w},p\right](i_{\tau})  K_{\tau}  - \sum_{\underline{i}\in S}i_{\tau}\frac{1}{n_c} \Bi\left[\frac{n_c}{w},px^{v-1}\right](i_{\tau})  K_{\tau}\right) \nonumber \\
&= px^{v-1}\frac{1}{w}\sum_{\tau\in[w]}\left( x - \sum_{\underline{i}\in \hat{S}} K_{\tau}\Bi[n_c/w-1,px^{v-1}](i_{\tau}) \right).
\label{eqn:local_convergence}
\end{align}
Note that the sum over degree types $\underline{i}\in \hat{S}\setminus S$ is due to a change of index variable. We do not require the solutions to differential equations of degree types not in $S$, hence the solutions in (\ref{eqn:solutions}) are sufficient in determining $e^T(x)$. 

The condition $e^T(x)>0$ is the necessary and sufficient convergence condition for the continuation of incremental decoding of the sub-ensemble at index $k$. This convergence condition is \emph{local} since we had fixed edge-bundles of types outside of the sub-ensemble. Note that specializing this condition, with $e^T(x)$ given by (\ref{eqn:local_convergence}), to the $(v,n_c/w)$-regular LDPC ensemble (with single parity-check component code $(n_c/w,n_c/w-1,2)$ and $w=1$), with the requirement that it holds for all $x\in(0,1]$, gives the convergence condition in \cite[Proposition 2]{luby:2001}. We continue to refer to this condition as a ``convergence'' condition, even though we use it to determine whether the incremental decoder can continue to decode, instead of convergence to the empty graph.

\subsection{Coupled ensemble}\label{subsec:coupled_ens}

To analyze the entire spatially-coupled split-component ensemble, we first expand the time horizon. Since in each sub-ensemble the expected number of initial edges is $Mn_cp$ and each iteration removes $v$ edges, the time steps in variable $l$ are $\{0,1,\dots,Mn_cp/v\}$. Over the entire coupled ensemble, the time steps are expanded to $\{0,1,\dots,Ln_cp/v\}$. Since we use the normalized time-variable $x$, this change is hidden. It is important to keep in mind that the time interval $(0,1]$ now represents an extended decoding process. 

In the following, each $x$ in the single-interleaver ensemble analysis, along with any associated functions (e.g. $e(x)$, $r^{k+\tau}_{(\underline{i})}(x)$), are distinguished in the spatial-dimension by the sub-script $k$. Since this becomes cumbersome for functions, we distinguish between functions at different spatial-locations by the sub-script of its \emph{variable} $x_k$. For example, the edge-count function for index $k$ is denoted by $e(x_k)$ and is distinct from $e(x_{k'})$ for all $k'\neq k$.

Consider the following decoding process over the coupled ensemble. At time $x_k=1$ for all $k\in[L]$, the sample graphs are generated and correctly received variable nodes and their edges are removed to obtain a collection of residual graphs. The average DDs are given by the initial conditions in (\ref{eqn:init_conds}). Applying incremental decoding in parallel at sub-ensembles at indices $k$ results in a complicated evolution of DDs with coupling between edge-bundles of different types.

If we assume that condition (\ref{eqn:frozen}) holds for each sub-ensemble at index $k$, the edge-bundles of different types are then effectively \emph{decoupled} at each DD. A partial description of the DD evolution, for degree types $\underline{i}\in S$, at each local decoder is given by (\ref{eqn:solutions}). Based on initial conditions and the local convergence condition (\ref{eqn:local_convergence}), we know (almost surely) when local decoding stops and the stopping DDs.

Define the \emph{stopping time} at index $k$ as
\[
x^s_k = \max_{x_k \in [0,1]} \{x_k | e^T(x_k)=0 \}.
\]
We refer to the progress of parallel incremental decoding (under assumption (\ref{eqn:frozen})) from an initial set of times (e.g. $x_k=1$) to stopping times $\{x^s_k\}_{k\in[L]}$, as a \emph{round} of decoding. We consider multiple rounds of decoding. Let $x^{s,q}_k$ denote the stopping times at the end of decoding round $q$, $q\geq 1$.

Stopping times may be different at different spatial locations. Since the functions $e(x_k)$, $e^T(x_k)$, and $r^{k+\tau}_{(\underline{i})}(x_k)$ no longer change for $x_k\leq x^s_k$ within a round of decoding, we define 
\[
x^{s,q} = \min_{k\in[L]}\{ x^{s,q}_k \}.
\]

The following proposition provides the bridge between single-interleaver and coupled analyses.

\begin{proposition}\label{prop:dd_prod}
For all $\underline{i}\in S$ and $k\in[L+w]$
\begin{equation}\label{eqn:updated_dd}
r^k_{(\underline{i})}(x^{s,q}) = \prod_{\tau\in[w]} \frac{1}{n_c}\Bi\left[ \frac{n_c}{w},p (x^{s,q}_{k-\tau})^{v-1} \right] (i_{\tau}) 
\end{equation}
\end{proposition}

\begin{IEEEproof}
The initial DDs are independent over edge-types as shown in (\ref{eqn:init_conds}). Given any degree type $\underline{i} \in S$ at index $k$ and stopping time $x^{s,q}_k$, each edge-bundle of type $\tau$ is independently modified by incremental decoding of the sub-ensemble at index $k-\tau$. Hence, the product form of the DD at index $k$ as shown in (\ref{eqn:init_conds}) is maintained throughout decoding round $q$. The overall DD for $\underline{i}\in S$ at index $k$ is given as
\begin{align*}
r^{k}_{(\underline{i})}(x^{s,q}) &= \prod_{\tau \in [w]} \sum_{\sim i_{\tau}} r^{k-\tau}_{(\underline{i})}(x^{s,q}_k) \\
&= \prod_{\tau \in [w]} \frac{1}{n_c} \Bi\left[ \frac{n_c}{w}, p (x^{s,q}_k)^{v-1} \right](i_{\tau}) \sum_{\sim i_{\tau}} K_{\tau} 
\end{align*}
\noindent
where we used (\ref{eqn:solutions}) with the notation $K_{\tau}$, evaluated at stopping time $x_k^{s,q}$. The sum in the last line evaluates to $1$ by the definition of $K_{\tau}$.

Note that we do not claim the independence of edge types within a degree type, even though the above expression appears to be a product of marginals. Clearly, after the first decoding iteration, the edge types within a degree type become dependent, since the condition for whether a constraint node is recoverable depends on the total degree of a degree type. 

The above expression gives the DD at index $k$ as a product of marginals obtained from \emph{different} DDs at $k-\tau$, $\tau\in[w]$ at the end of a decoding round. The product form is a consequence of the deterministic splitting of constraint nodes and does not require independence.
\end{IEEEproof}

We represent the product of binomial probability mass functions in (\ref{eqn:updated_dd}) by the vector $\underline{x}^q_k=(x^{s,q}_{k},x^{s,q}_{k-1},\dots,x^{s,q}_{k-w+1})$. More precisely, given $\underline{x}^q_k$ we evaluate the functional 
\[
\mathcal{B}(\underline{x}^q_k) =  \prod_{\tau\in[w]} \Bi[w/n_c,p(x_{k-\tau}^{s,q})^{v-1}](i_{\tau})
\]
to obtain the function $r^k_{(\underline{i})}(x^{s,q})$.

At the end of decoding round $q$, we update the DDs based on (\ref{eqn:updated_dd}). The updated DDs then become the initial conditions $r^k_{(\underline{i})}(x^{s,q})$ for decoding round $q+1$. An additional round of decoding is possible since the updated DDs may once again allow the local convergence condition (\ref{eqn:local_convergence}) to be satisfied at some spatial locations.

\begin{proposition}\label{prop:global_convergence}
A necessary and sufficient condition for the almost-sure convergence of a residual graph from the $\mathcal{E}((n_c,k_c,d_c),v,M,w)$ spatially-coupled split-component ensemble to the empty graph, with transmission over a BEC with erasure probability $p$, as $M\rightarrow\infty$ is
\begin{equation}\label{eqn:global_convergence}
\frac{1}{w}\sum_{\tau\in[w]} \sum_{\underline{i} \in \hat{S}} \prod_{\gamma\in[w]}
\Bi[n_c/w-\mathbb{I}[\gamma=\tau],p(x_{k+\tau-\gamma}^{s,q})^{v-1}](i_{\tau}) < x_k^{s,q}
\end{equation} 
for all $k\in[L]$, where $x^{s,q}_k$ are given by the recursion
\begin{multline*}
x^{s,q+1}_k = \\
\max_{x_k\in(0,1]}\left\{x_k | x_k - \sum_{\underline{i}\in \hat{S}} K_{\tau}\Bi[n_c/w-1,p(x^{s,q}_kx_k)^{v-1}](i_{\tau})>0 \right\}.
\end{multline*}
\noindent
Under initial conditions
\[
x^{s,0}_{k+\tau-\gamma} = \mathbb{I}[0\leq k+\tau-\gamma<L] 
\] 
and boundary conditions
\[
x^{s,q}_{k+\tau-\gamma} = 0\; \textrm{ if } k+\tau-\gamma \notin [L].
\]
\end{proposition}

\begin{IEEEproof}
At the end of decoding round $q$, all expected DD evolutions have stopped at stopping time $x^{s,q}_k$ for each index $k$. This is due to the violation of the local convergence condition (\ref{eqn:local_convergence}) under assumption (\ref{eqn:frozen}).

From Proposition \ref{prop:dd_prod}, the \emph{actual} DDs at the end of round $q$, without fixing any edge-bundles, are given by (\ref{eqn:updated_dd}) for all $\underline{i}\in S$. Update the local convergence condition (\ref{eqn:local_convergence}) using (\ref{eqn:updated_dd}). Note that $e^T(x)$ is now a function of $(x^{s,q}_k,x^{s,q}_{k+1},\dots,x^{s,q}_{k+w-1})$. Simplifying the condition $e^T(x^{s,q}_k,x^{s,q}_{k+1},\dots,x^{s,q}_{k+w-1})>0$ gives (\ref{eqn:global_convergence}). 

This condition only allows the decoder to continue and begin decoding round $q+1$. At the beginning of round $q+1$, the initial conditions for the DD differential equations are $p(x^{s,q}_k)^{v-1}$, for the sub-ensemble at index $k$. Hence, the solutions are given by (\ref{eqn:solutions}) with $p$ replaced by $p(x^{s,q}_k)^{v-1}$. Using this solution in the stopping time definition gives the recursion. The initial and boundary conditions correspond to the average initial DDs and suppressed edge-bundles in the ensemble definition.
\end{IEEEproof}

This \emph{global} convergence condition can be simplified by using the Poisson distribution $\Po[np](i)$ to approximate the binomial distribution $\Bi[n,p](i)$. While the approximation is exact only for $n_c\rightarrow\infty$, the differences are negligible for $n_c/w>20$ and $p<0.05$. The Poisson approximation was also used in the analysis of \cite{justesen:2011} for iterative algebraic component-code decoding of product codes.

Define the column vector $\mathbf{x}^q=(x^{s,q}_0,\dots,x^{s,q}_{L-1})^T$ and the function 
\[
\pi(a,n,x) =\sum_{i:i\geq a}\Po[nx](i).
\]
Let
\begin{equation}
f(x,p) = p x^{v-1}  \text{ and }
g(x) = \pi(d_c-1,n_c,x).  \label{eqn:singleRecursion}
\end{equation}
We use $\mathbf{f}(\mathbf{x},p)$ and $\mathbf{g}(\mathbf{x})$ to denote
the vector obtained by the evaluation of $f$ and $g$ at the components
of vector $\mathbf{x}$. Let $A$ be a $L \times (L+w-1)$ spatial-coupling matrix with entries given by
$A_{i,j} = \frac{1}{w} \cdot \mathbb{I}[ j-i\in[w] ]$.

\begin{theorem}\label{thm:coupledRecursion}
For any sample graph from a spatially-coupled split-component ensemble $\mathcal{E}((n_c,k_c,d_c),v,M,w)$ with $M\rightarrow\infty$, transmitted over a BEC with erasure probability $p$. Under decoding by Algorithm \ref{alg:peelDecWithVar}, the fixed points of the decoding process are equivalent to the fixed points of the recursion

\begin{equation}\label{eqn:v2_recursion}
\mathbf{y}^q=A^T\mathbf{f}(A\mathbf{g}(\mathbf{y}^{q-1}),p)
\end{equation}
where 
\[
\mathbf{y}^q = A^T \mathbf{f}(\mathbf{x}^q,p)
\]
and
\[
\mathbf{x}^q = A \mathbf{g}(\mathbf{y}^q)
\]
with initial and boundary conditions on $x^q_k$ given in Proposition \ref{prop:global_convergence}.
\end{theorem}

\begin{IEEEproof}
Using the Poisson approximation, the sum in (\ref{eqn:global_convergence}) can be written as
\begin{multline*}
\frac{1}{w}\sum_{\tau\in[w]} \sum_{\underline{i} \in \hat{S}} \prod_{\gamma\in[w]}
\Po\left[\frac{pn_c}{w}(x_{k+\tau-\gamma}^{s,q})^{v-1}\right](i_{\tau}) \\
= \frac{1}{w}\sum_{\tau\in[w]}\sum_{s:s\geq d_c-1}\Po\left[\frac{1}{w}\sum_{\gamma\in[w]}pn_c(x_{k+\tau-\gamma}^{s,q})^{v-1}\right](s)\\
=\frac{1}{w}\sum_{\tau\in[w]}\pi\left(d_c-1,n_c,\frac{1}{w}\sum_{\gamma\in[w]}p(x^{s,q}_{k+\tau-\gamma})^{v-1}\right).
\end{multline*} 
This is equivalent to
\[
A \mathbf{g}( A^T\mathbf{f}(\mathbf{x}^q,p) ).
\]
Therefore all $\mathbf{x}$ which satisfy 
\[
\mathbf{x} = A \mathbf{g}( A^T\mathbf{f}(\mathbf{x},p) )
\] 
are also sets of stopping times at which coupled decoding must stop. 

Conversely, since (\ref{eqn:global_convergence}) are the normalized number of recoverable edges, it is non-negative. Therefore, when the conditions are violated and the decoder stops, (\ref{eqn:global_convergence}) must hold with equality for all $k$, which is equivalent to a fixed point of the above recursion. 

Since the recursion $\mathbf{y}^q$ is a ``half-iteration shift'' of the recursion $\mathbf{x}^q$, they have the same fixed points. 
\end{IEEEproof}

\subsection{Potential threshold}

Verifying the conditions in \cite[Def. 34]{pfister:2014}, we conclude that the recursion in Theorem \ref{thm:coupledRecursion} is an unconditionally-stable admissible system. Hence, we can apply the potential function analysis of \cite{pfister:2014} to 
determine the \emph{potential threshold} \cite[Def. 35]{pfister:2014} of spatially-coupled split-component ensembles for the BEC. The most relevant details of the potential function analysis are briefly outlined in Appendix \ref{append:potFcn}. 

The single-system potential function for the admissible system of Theorem \ref{thm:coupledRecursion} is given by
\begin{align}
U_s(x,p) &= x\pi(d_c-1,n_c,x) - \left(x-\frac{1}{n_c}\sum_{i=0}^{d_c-2}\pi(i+1,n_c,x)\right) \nonumber \\
&- p\frac{1}{v}\pi(d_c-1,n_c,x)^v \label{eqn:pot_fcn}.
\end{align}
The fixed-point potential is
\begin{align}
Q(x) &= \left(1-\frac{1}{v}\right)x\pi(d_c-1,n_c,x) - \left(x-\frac{1}{n_c}\sum_{i=0}^{d_c-2}\pi(i+1,n_c,x)\right) \nonumber \\
&= \left(\frac{d_c-1}{n_c}-\frac{x}{v}\right)\pi(d_c-1,n_c,x) - \frac{d_c-1}{n_c}\Po[n_cx](d_c-1). \label{eqn:splitRecurFixedPointPot}
\end{align}
\noindent
The potential threshold can then be obtained by using Lemma \ref{lemma:fp_pot}.

\begin{figure}
\centering
\includegraphics[scale=0.85]{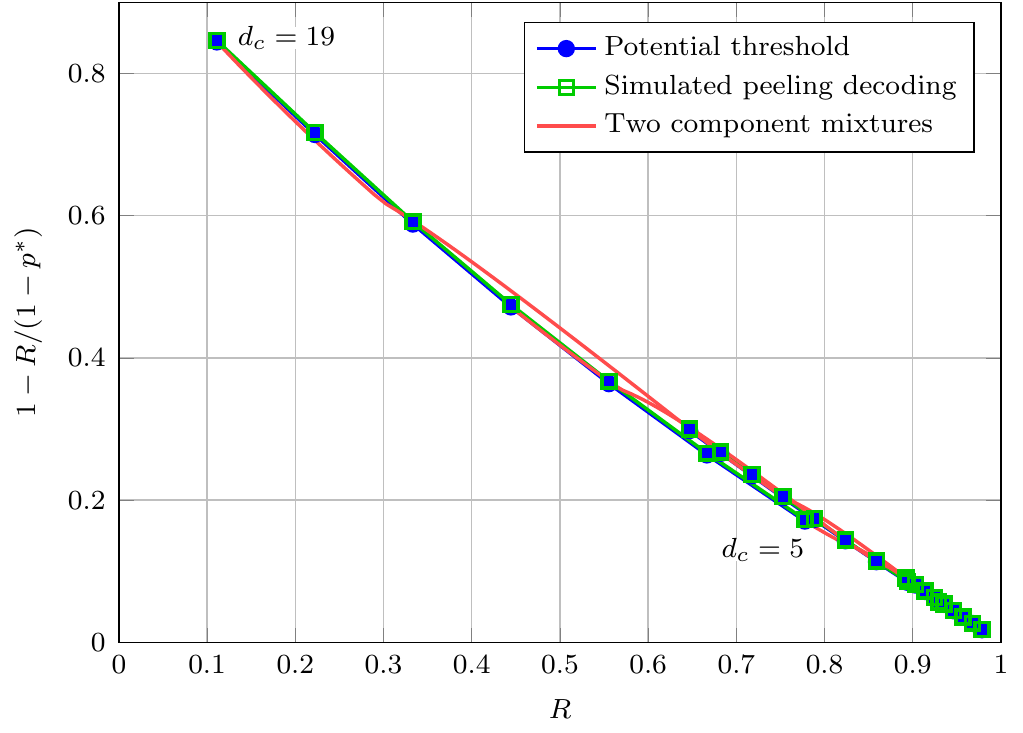}
\caption{Multiplicative gaps to capacity of analytical and simulated thresholds of $\mathcal{E}(\mathcal{C},2,M,w)$ spatially-coupled split-component ensemble over the BEC. Component codes are shortened BCH codes with $n_c=2^{m_c}-2$, $m_c\in\{7,9,11\}$, and $d_c\in\{5,7,\dots,21\}$. Code graphs for simulations were sampled uniformly at random from the $\mathcal{E}(\mathcal{C},2, \lfloor 2\times10^6/n_c \rfloor,2)$ ensemble and simulated thresholds were based on an output erasure probability of $10^{-6}$. Only codes with positive rates are shown. Analytical thresholds of mixtures of two BCH component codes are also shown.}
\label{fig:BCHv2PotThresWithSimBEC}
\end{figure}

Figure \ref{fig:BCHv2PotThresWithSimBEC} shows the analytical and simulated thresholds of spatially-coupled split-component ensemble $\mathcal{E}(\mathcal{C},2,M,w)$ based on shortened BCH component codes, in terms of their \emph{multiplicative gaps to capacity}. Given the threshold channel parameter $p^*$ of an ensemble or code of rate $R$, we define the multiplicative gap to capacity to be the $\epsilon$ such that $R=(1-\epsilon)C(p^*)$, where $C(p^*)$ is the capacity of the channel with parameter $p^*$. Observe that the potential thresholds of $\mathcal{E}(\mathcal{C},2,M,w)$ ensembles under the BEC approaches capacity as $R\rightarrow 1$ and decreases with decreasing rate.

\subsection{Weight-pulling threshold}

We now present an intuitive upper bound on the number of
erasures that a code in our ensemble can recover.
Each local decoder can recover at most $d_c-1$ erasures.  
Since there are $M(L+w-1)$ constraint nodes
in the graph, the code can recover
at most $M(L+w-1)(d_c-1)$ erasures, and this only if each
decoder is operating at its maximum erasure-recovering capability,
i.e., each decoder ``pulls its own weight.''
Note that a total of $NL$ bits are transmitted.
This motivates
the following definition.

\begin{definition}\label{def:weightPulling}
Define the \emph{weight-pulling threshold} for a $\mathcal{E}(\mathcal{C},v,M,w)$ ensemble as
\begin{equation*}
p^*_{\textrm{wp}}= \lim_{L \rightarrow \infty} \frac{(d_c-1)M(L+w-1)}{NL}=v \frac{d_c-1}{n_c}.
\end{equation*}
\end{definition}

\begin{proposition}\label{prop:convPotThres}
The potential threshold of the $\mathcal{E}(\mathcal{C},v,M,w)$ ensemble converges to $p^*_{\textrm{wp}}$ as $d_c\rightarrow\infty$. 
\end{proposition}
\begin{IEEEproof}
Let $x_0=v(d_c-1)/n_c$. By the trivial bound $d_c\leq n_c$, we have $x_0<\infty$ for any $d_c$. From (\ref{eqn:splitRecurFixedPointPot}), we conclude that $x_0$ is a root of $Q(x)$ for $d_c\rightarrow\infty$. From the second derivative of $Q(x)$
\begin{equation*}
\frac{d^2}{dx^2}Q(x)= \Po[n_cx](d_c-2) \left[ \left(1-\frac{2}{v}\right)n_c + \left(1-\frac{1}{v}\right)(d_c-2)n_c - \left(1-\frac{1}{v}\right)n_c^2 x\right] 
\end{equation*}
\noindent
we see there is a unique inflection point at
\[
x'=\frac{1}{n_c}\left(d_c-1-\frac{1}{v-1}\right)
\]
with $Q(x)$ convex for $x\in[0,x']$ and concave for $x\in[x',x_{\textrm{max}}]$. Since $v\geq 2$ and $Q(0)=0$, $x_0$ is the unique non-trivial root of $Q(x)$. By Lemma \ref{lemma:fp_pot}, the potential threshold is given by $\tilde{\lambda}(x_0)=x_0/[\pi(d_c-1,n_c,x_0)]^{v-1}=v(d_c-1)/n_c$ as $d_c\rightarrow\infty$.
\end{IEEEproof}
 		
%
%

\section{Mixture ensembles}\label{sec:genEnsem}

In this section, we define and analyze spatially-coupled
split-component ensembles made up of mixtures of component codes. Let
$\mathcal{T}=\{(n_{i,c},k_{i,c},d_{i,c})\}_{i=0}^{T-1}$
be a collection of binary
linear block codes. For the set of non-negative integers
$\mathcal{M}=\{M_0,\dots,M_{T-1}\}$, let
$\mathcal{E}(\mathcal{T},v,\mathcal{M},w)$ denote the \emph{mixture ensemble}
containing $M_i$ component codes of type $(n_{i,c},k_{i,c},d_{i,c})$ at each
spatial-location, with variable degree $v$ and coupling width $w$. Each component code is split uniformly into $w$ edge-types. The edge balance
condition is $N=\sum_{i\in[T]}M_{i,c}n_{i,c}/v$. 
All other ensemble definitions remain the same.

The \emph{mixture distribution} of a $\mathcal{E}(\mathcal{T},v,\mathcal{M},w)$ ensemble is the set 
$\rho=\{\rho_0,\dots,\rho_{T-1}\}$ with
\begin{equation*}
\rho_i=\frac{{M_i}n_{i,c}}{\sum_{i'\in[T]}M_{i'}n_{i',c}}.
\end{equation*}
The design rate of a mixture ensemble is given by 
\[
R = 1 - v \left( 1-\sum_{i\in[T]}\rho_iR_{i,c} \right).
\]
\noindent
The same initialization, decoding algorithm, and definition of $\mathbf{y}^q$ in Sec.~\ref{sec:potThresStairEns} for spatially-coupled split-component ensembles apply to the mixture ensemble. For convenience, let 
\[
\pi_{\rho}(x)=\sum_{i\in[T]}\rho_i\pi(d_{i,c}-1,n_{i,c},x).
\]

\begin{theorem}\label{thm:coupledRecursionMixEnsem}
For any sample graph from a mixture ensemble $\mathcal{E}(\mathcal{T},v,\mathcal{M},w)$ with $\sum_i M_i\rightarrow\infty$, transmitted over a BEC with erasure probability $p$. Under decoding by Algorithm \ref{alg:peelDecWithVar}, the fixed points of the decoding process are equivalent (almost surely) to the fixed points of the recursion in Theorem \ref{thm:coupledRecursion} with functions $f,g$ defined as
\[
f(x,p) = px^{v-1} \text{ and } g(x)=  \pi_{\rho}(x).
\]
\end{theorem}

\begin{IEEEproof}
Straightforward modification of the derivation in Sec. \ref{sec:potThresStairEns} and proof of Theorem \ref{thm:coupledRecursion}, with an additional sum over the set of component codes $\mathcal{T}$ with weights $\rho$. This modification directly leads to the new definition of $g(x)$ in the theorem statement.
\end{IEEEproof}

The single-system potential function and fixed-point potential for the recursion in Theorem \ref{thm:coupledRecursionMixEnsem} are
\begin{align}
U_s(x,p) &= x \pi_{\rho}(x) \nonumber \\
&- \sum_{i\in[T]}\rho_i\left(x-\frac{1}{n_{i,c}}\sum_{j=0}^{d_{i,c}-2}\pi(d_{i,c}-1,n_{i,c},x)\right) \nonumber \\
&- p\frac{1}{v}[\pi_{\rho}(x)]^v \label{eqn:potFuncMixEnsem}
\end{align}
\noindent
and
\begin{align}
Q(x) &= \left(1-\frac{1}{v}\right)x\pi_{\rho}(x) \nonumber \\
&- \left(x - \sum_{i\in[T]}\frac{\rho_i}{n_{i,c}}\sum_{j=0}^{d_{i,c}-2}\pi(d_{i,c}-1,n_{i,c},x)\right). \label{eqn:FixedPointPotMixEnsem}
\end{align}

In Fig. \ref{fig:BCHv2PotThresWithSimBEC} we also plot the analytical thresholds of $\mathcal{E}(\mathcal{T},2,\mathcal{M},w)$ mixture ensembles with $\mathcal{T}$ containing two shortened BCH component codes with block-lengths $n_{c,1}=2^{m_c}-2$, $n_{c,2}=2^{m_c+2}-2$ for $m_c\in\{7,9\}$ and  the same $d_c\in\{5,11,19\}$.
Observe that the potential thresholds of mixture ensembles smoothly interpolate
between the potential thresholds of single-component spatially-coupled split-component ensembles. Hence, the mixture ensemble allows for a greater range of rates to be achieved than shortening of a single component code. Furthermore, using pairs of component codes with the same $d_c$, a small threshold increase for rates \emph{between} the rates of single-component ensembles can be obtained. 

\begin{definition}
Define the weight-pulling threshold of a $\mathcal{E}(\mathcal{T},v,\mathcal{M},w)$ mixture ensemble as
\begin{equation*}
p_{\textrm{wp}}^* = \lim_{L\rightarrow \infty} \frac{\sum_i M_i(d_{i,c}-1)(L+w-1)}{NL} =
v\sum_i\rho_i\frac{d_{i,c}-1}{n_{i,c}}.
\end{equation*}
\end{definition}

\begin{proposition}\label{prop:convMixPotThres}
The potential threshold of the $\mathcal{E}(\mathcal{T},v,\mathcal{M},w)$
ensemble where $d_{i,c}=d_c$ for all $i\in[T]$ converges to $p^*_{\textrm{wp}}$
as $d_c\rightarrow\infty$.
\end{proposition}

\begin{IEEEproof}
Let $x_0=v\sum_i\rho_i(d_c-1)/n_{i,c}$.
We re-write (\ref{eqn:FixedPointPotMixEnsem}) as 
\begin{equation*}
Q(x) = \left[\frac{\sum_i\rho_i(d_c-1)/n_{i,c}}{\pi_{\rho}(x)}-\frac{\sum_i\rho_i (d_c-1)/n_{i,c}\sum_{j=0}^{d_c-1}\Po[n_{i,c}x](j)}{\pi_{\rho}(x)} - \frac{x}{v}\right]\pi_{\rho}(x).
\end{equation*}
\noindent
Evaluating $Q(x_0)$, we see that since $\pi_{\rho}(x_0)\rightarrow 1$ and the second term in the sum tends to 0 as $d_c\rightarrow \infty$, $x_0$ is a root of $Q(x)$. Since $d_{i,c}=d_c$ for all $i\in[T]$ and (\ref{eqn:FixedPointPotMixEnsem}) is a convex combination of (\ref{eqn:splitRecurFixedPointPot}), the convexity argument in the proof of Proposition \ref{prop:convPotThres} also holds in this case. Hence, $x_0$ is the unique root. The potential threshold is given by $\tilde{\lambda}(x_0)=x_0/[\pi_{\rho}(x_0)]^{v-1}=v\sum_i\rho_i(d_c-1)/n_{i,c}$ as $d_c\rightarrow\infty$.
\end{IEEEproof}
 		
%
%

\section{BSC analysis of spatially-coupled split-component ensembles}\label{sec:bscApproximation}

The analysis of Sec.~\ref{sec:potThresStairEns} for the BEC can be used to approximate the potential threshold of spatially-coupled split-component ensembles for the BSC. Let $t_c=\lfloor (d_c-1)/2 \rfloor$ denote the \emph{unique decoding radius} of a $(n_c,k_c,d_c)$ binary linear code. We refer to a component-code decoder as \emph{mis-correction-free} if it makes corrections when $\leq t_c$ bit errors are present in the received word and declares failure otherwise. In the following, we assume that all component-code decoders are mis-correction-free. It is known that the mis-correction probability of Reed-Solomon codes is proportional to $1/t_c!$ \cite{justesen:2011,jian:2012}. Hence, we expect our analysis to be inaccurate for small values of $t_c$ and to improve quickly with increasing $t_c$. 

The BEC analysis in Theorem \ref{thm:coupledRecursion} is modified to the mis-correction-free BSC case by replacing $g(x)=\pi(d_c-1,n_c,x)$ by $g(x)=\pi(t_c,n_c,x)$ in (\ref{eqn:singleRecursion}) and defining the BEC erasure probability $p$ to be the BSC bit-error probability. The potential threshold for the modified vector recursion is then the potential threshold for the BSC. The weight-pulling threshold for the BEC can also be modified to the BSC case by replacing $d_c-1$ in Definition \ref{def:weightPulling} by $t_c$ to obtain
\[
p^*_{\textrm{wp}} = v\frac{t_c}{n_c}.
\]

\begin{figure}
\centering
\includegraphics[scale=0.85]{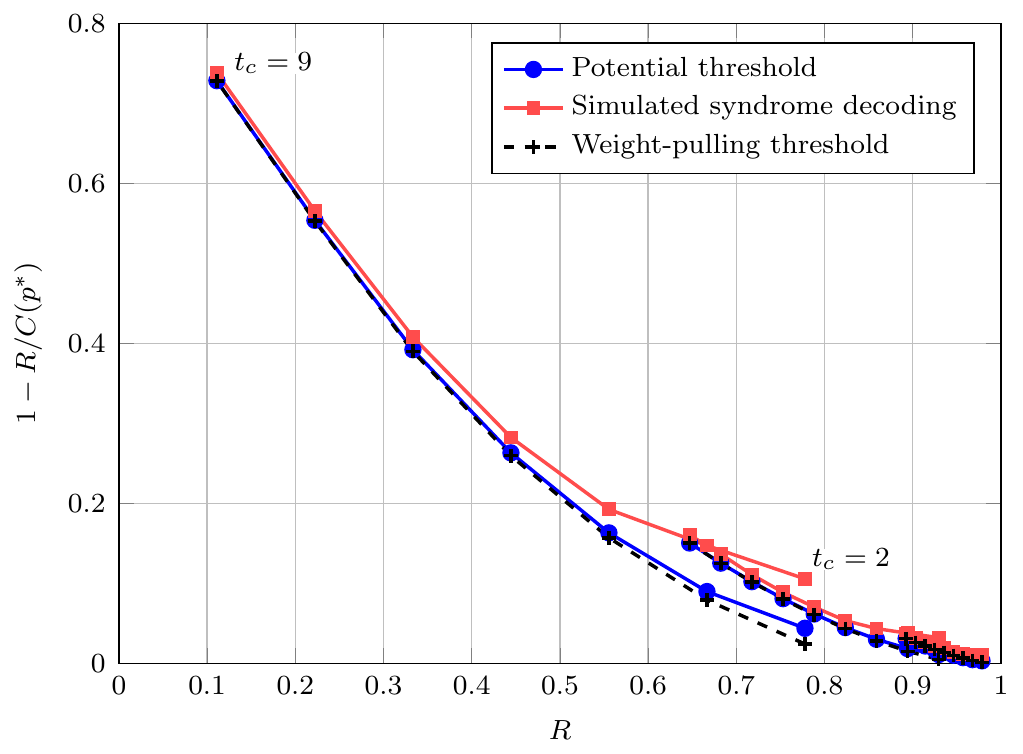}
\caption{Multiplicative gaps to capacity of analytical (under mis-correction-free decoding) and simulated thresholds of $\mathcal{E}(\mathcal{C},2,M,w)$ spatially-coupled split-component ensembles over the BSC. Component codes are shortened BCH codes with $n_c=2^{m_c}-2$, $m_c\in\{7,9,11\}$, and $d_c\in\{5,7,\dots,19\}$. Code graphs for simulations were sampled uniformly at random from the $\mathcal{E}(\mathcal{C},2, \lfloor 2\times10^6/n_c \rfloor,2)$ ensemble and simulated thresholds were based on an output error probability of $10^{-6}$. Only codes with positive rates are shown.}
\label{fig:BCHv2PotThresWithSimBSC}
\end{figure}

Figure \ref{fig:BCHv2PotThresWithSimBSC} shows the analytical and simulated thresholds of spatially-coupled split-component ensemble $\mathcal{E}(\mathcal{C},2,M,w)$ based on shortened BCH component codes over the BSC, under the mis-correction-free decoding assumption. The analysis appears to be inaccurate for small values of $t_c$ and improves as $t_c$ increases. Furthermore, since by Proposition \ref{prop:convPotThres} the potential threshold saturates to the weight-pulling threshold as $t_c\rightarrow\infty$, the threshold of spatially-coupled split-component ensemble \emph{with mis-corrections} also saturates to the weight-pulling threshold for large $t_c$. This is confirmed by the overlays in Fig. \ref{fig:BCHv2PotThresWithSimBSC}. 

%
%

\section{Beyond bounded-distance decoder}\label{sec:genCompDec}

The potential and weight-pulling thresholds for the BSC suggest that spatially-coupled split-component ensembles are bounded away from capacity, even if component-code decoders are mis-correction-free. 
The key reason is that the bounded-distance decoding of component codes neglects the full error-correcting capability of each component code. In this section, we consider theoretical potential threshold improvements assuming component-code decoders which can correct a fraction of errors \emph{beyond} the unique decoding radius.

For any $(n_c,k_c,d_c)$ binary linear code and decoder, let
$\beta=\{0,\dots,0,\beta_{t_c+1},\dots,\beta_{t_m},1,1,\dots\}$ be the
\emph{decoding profile}, where $\beta_i$ is the fraction of weight $i$ error
patterns the decoder \emph{cannot} correct. Under the mis-correction-free
assumption, we modify Algorithm~\ref{alg:peelDecWithVar} as follows: at each
iteration $l$, a constraint node with total degree $i$
is deemed correctable with probability $1-\beta_i$. We define the function 
\[
\pi_{\beta}(a,n,x) \triangleq \sum_{i:i\geq a}\beta_{i+1}\textrm{Po}[nx](i).
\] 

\begin{theorem}\label{thm:coupledRecursionBeyondBBDSpatialEnsem}
For any sample graph from a spatially-coupled split-component ensemble $\mathcal{E}(\mathcal{C},v,M,w)$ with $M\rightarrow\infty$, transmitted over a BSC with bit-error probability $p$. Under decoding by Algorithm \ref{alg:peelDecWithVar} with component decoders having decoding profile $\beta$, for all $w\geq w_0$ where $w_0$ only depends on $\beta$, an upper-bound on the recursion (almost surely) describing the decoding process is given by the recursion in Theorem \ref{thm:coupledRecursion}, with functions $f,g$ defined as
\[
f(x,p) = px^{v-1} \text{ and } g(x) = \pi_{\beta}(t_c,n_c,x).
\]
\end{theorem}
\begin{IEEEproof}

Following the steps given in Sec. \ref{sec:potThresStairEns}, we only modify the arguments for constraint nodes of total degree in the set $\{t_c+1,\dots,t_m\}$, which we refer to as \emph{fractionally-recoverable} constraint nodes. We generalize the sets $T = \{ \underline{i} | \sum_{\tau\in[w]} i_{\tau} \leq t_m \}$ and $S = \{ \underline{i} | \sum_{\tau\in[w]} i_{\tau} \geq t_c+1 \}$, which now have non-empty intersection. Since the decoding profile only applies to the total degree of a constraint node, given degree type $\underline{i}$, we denote its total degree by $u(\underline{i})=\sum_{\tau \in [w]}i_{\tau}$.

At the end of the first round of decoding, i.e. from initialization to the first set of stopping times, at each spatial-location a fraction of $(1-\beta_{u(\underline{i})})$ fractionally-recoverable constraint nodes of total degree $u(\underline{i})$ have been removed, leaving a fraction of $\beta_{u(\underline{i})}$. A straightforward extension of the derivation of (\ref{eqn:solutions}) gives the differential equation solutions
\begin{align}
r^{k+\tau}_{(\underline{i})}(x) &= \frac{1}{n_c} \beta_{u(\underline{i})} \Bi\left[\frac{n_c}{w},px^{v-1}\right](i_{\tau}) \prod_{\gamma\in [w] \setminus \tau} \Bi\left[n_c/w,p \cdot \mathbb{I}[0 \leq k + \tau - \gamma < L]\right](i_{\gamma}) \nonumber \\
&\triangleq \frac{1}{n_c} \beta_{u(\underline{i})} \Bi\left[\frac{n_c}{w},px^{v-1}\right](i_{\tau}) \prod_{\gamma\in [w] \setminus \tau} K(i_{\gamma}) \label{eqn:bbd_solutions}
\end{align}
for all $\underline{i}\in S$. Here we write the constant $K_{\tau}$ in a more explicit way by using $K(i_{\gamma})$ (c.f. (\ref{eqn:solutions}) and discussions therein).

Given decoding profile $\beta$, Proposition \ref{prop:dd_prod} is modified to
\begin{align}
r^{k}_{(\underline{i})}(x^{s,q}) &= \left( \prod_{\tau \in [w]} \frac{1}{n_c} \Bi\left[ \frac{n_c}{w}, p (x^{s,q}_k)^{v-1} \right](i_{\tau}) \right) 
\left( \prod_{\tau \in [w]} \sum_{s \geq t_c+1} \beta_s \sum_{\underline{i}':u(\underline{i}')=s}\prod_{\gamma\in[w]\setminus \tau} K(i'_{\gamma}) \right) \label{eqn:bbd_stopping_dd_line1}\\
&\leq \beta_{u(\underline{i})} \prod_{\tau \in [w]} \frac{1}{n_c} \Bi\left[ \frac{n_c}{w}, p (x^{s,q}_k)^{v-1} \right](i_{\tau}) \label{eqn:bbd_stopping_dd}
\end{align}
for all $\underline{i}\in S$, $k\in[L+w]$, and $w\geq w_0$ for some positive integer $w_0<\infty$. The inequality is obtained by first recognizing that for bounded-distance decoding without mis-corrections, where $\beta_s=1$ for all $t_c < s \leq n_c$ and $\beta_s=0$ otherwise, the rightmost factor in (\ref{eqn:bbd_stopping_dd_line1}) is equal to 1, identical to the derivation of Proposition \ref{prop:dd_prod}. For any other decoding profile, the rightmost factor in (\ref{eqn:bbd_stopping_dd_line1}) is upper-bounded by a non-negative constant less than 1 raised to the $w$th power. Hence, there must exist a $w_0$ such that $\prod_{\tau \in [w]} \sum_{s \geq t_c+1} \beta_s \sum_{\underline{i}':u(\underline{i}')=s}\prod_{\gamma\in[w]\setminus \tau} K(i'_{\gamma}) \leq \beta_{u(\underline{i})}$.

Finally, using the Poisson approximation and (\ref{eqn:bbd_stopping_dd}), we obtain an upper-bound on the recursion expression (\ref{eqn:global_convergence})
\begin{align*}
\frac{1}{w}\sum_{\tau\in[w]} \sum_{\underline{i} \in \hat{S}} \beta_{u(\underline{i})} \prod_{\gamma\in[w]}
\Po\left[\frac{pn_c}{w}(x_{k+\tau-\gamma}^{s,q})^{v-1}\right](i_{\tau}) \\
&= \frac{1}{w}\sum_{\tau\in[w]}\sum_{s:s\geq t_c} \beta_s \Po\left[\frac{1}{w}\sum_{\gamma\in[w]}pn_c(x_{k+\tau-\gamma}^{s,q})^{v-1}\right](s)\\
&= \frac{1}{w}\sum_{\tau\in[w]}\pi_{\beta}\left(t_c,n_c,\frac{1}{w}\sum_{\gamma\in[w]}p(x^{s,q}_{k+\tau-\gamma})^{v-1}\right).
\end{align*} 

\end{IEEEproof}

\begin{theorem}[\cite{helleseth:2005}]\label{thm:boundsDecProf}
For any $(n,k,d)$ binary linear code with unique decoding radius $t$, for
$i\in\{t+1,\dots,t_m\}$ its decoding profile satisfies
$1-\beta_i\leq\sigma(n,k,i)^{-1}$ where
$\sigma(n,k,i)=2^{k-n}\sum_{j=0}^{i}\binom{n}{j}$. Moreover, there exists a
code such that $\beta_i\leq\sigma(n,k,i)$.
\end{theorem}

The potential function analysis can be applied to find the potential threshold of the coupled recursion in Theorem \ref{thm:coupledRecursionBeyondBBDSpatialEnsem}, which is an upper-bound on the potential threshold of beyond bounded-distance decoder. One detail in the potential function analysis is that the potential threshold only applies for $w>w_1$ where $w_1<\infty$ is a positive integer (see Appendix \ref{append:potFcn}). Here, we take $w_1=\max(w_0,w_1)$ where $w_0$ was given in Theorem \ref{thm:coupledRecursionBeyondBBDSpatialEnsem}.

\begin{figure}
\centering
\includegraphics[scale=0.85]{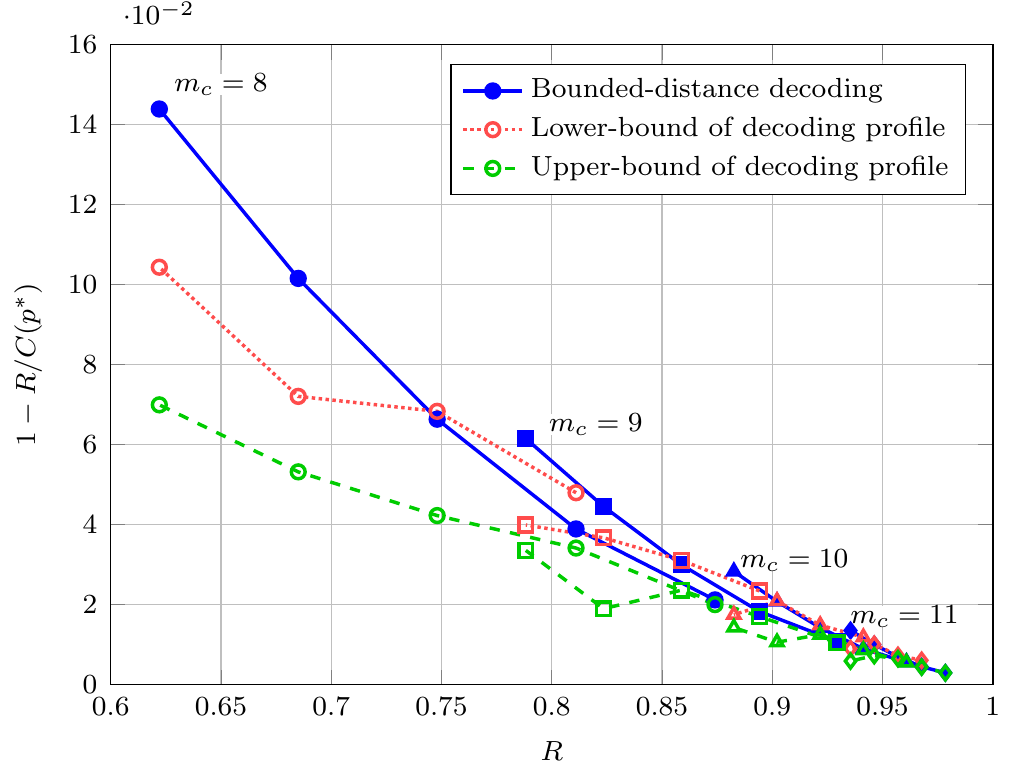}
\caption{Multiplicative gaps to capacity of analytical thresholds of $\mathcal{E}(\mathcal{C},2,M,w)$ ensembles under bounded-distance decoding and beyond bounded-distance decoding with decoding profiles satisfying existence lower-bounds and theoretical upper-bounds given in Theorem \ref{thm:boundsDecProf}. The component codes are primitive BCH codes with $m_c\in\{8,9,10,11\}$, where different values of $m_c$ are distinguished by different markers. For each $m_c$, potential thresholds of ensembles with BCH component codes $(2^{m_c}-1,2^{m_c}-1-m_ct_c,2t_c+1)$ for $t_c\in\{2,3,4,5,6\}$ are plotted. Thresholds corresponding to different component decoders are distinguished by line styles.}
\label{fig:cosetBoundsPotThres}
\end{figure}

Figure \ref{fig:cosetBoundsPotThres} shows the analytical thresholds of $\mathcal{E}(\mathcal{C},2,M,w)$ ensembles under beyond bounded-distance decoding, based on the existence lower-bound and theoretical upper-bound on the decoding profile given in Theorem \ref{thm:boundsDecProf}. The results suggest that significant improvements are possible, since thresholds calculated from existence lower-bounds of decoding profiles exceed thresholds based on bounded-distance decoding of the same component code. However, whether such error-correction capabilities can be achieved with an \emph{efficient} decoder is an open problem.

%
%
%

\section{Conclusions and Future Work}

In this paper, we proposed the spatially-coupled split-component ensemble, a generalization of the code structures in a number of efficiently-decodable error-correction coding schemes for high bit-rate optical-fiber communication systems. In our definition, the deterministic splitting of the edges of constraint nodes was preserved, as was the efficient intrinsic hard-decision iterative algebraic decoding algorithm. We analyzed the threshold of spatially-coupled split-component ensembles over the BEC by using a generalized peeling decoder and the differential equation method. A vector recursion describing the expected behavior of the ensemble throughout the decoding process was derived and analyzed using known techniques. We showed the convergence of the threshold to the upper-bound $v(d_c-1)/n_c$ as the minimum distance of the component code increased. 

We defined the mixture ensemble, in which a number of different component codes were present at each spatial location, and studied their thresholds. By assuming mis-correction-free decoding, we used the BEC analysis to approximate the BSC thresholds of spatially-coupled split-component ensembles. We demonstrated that the effects of mis-correction were significant for $t_c\leq 4$ and became negligible as $t_c$ increases. One of the main conclusions of this work is that spatial-coupling achieves the theoretical upper-bound for bounded-distance decoding, $vt_c/n_c$, for large values of $t_c$. Finally, we studied the thresholds of spatial-coupled split-component ensembles over the BSC, assuming component decoders could correct some fraction of error-patterns of weights beyond $t_c$. Our results indicate that thresholds may be improved by using such decoders, however the complexity cost is likely significant.

An immediate topic of future work is to extend the analysis to the finite-length regime where $M<\infty$. The generalized peeling decoding studied in this paper can be extended to the finite-length regime by considering the covariance evolution around some critical parameter (e.g. the fraction of recoverable constraint nodes at each spatial-location) throughout the decoding process. A similar technique was used for the finite block-length analysis of spatially-coupled LDPC ensembles over the BEC \cite{olmos:2015}, where the critical parameter was the fraction of degree-one check nodes in the entire spatially-coupled chain.

Another topic of future work is to study possible threshold improvements when list decoders are used to decode a fraction of the constraint nodes. We may consider a mixture of BCH and Reed-Solomon (RS) component codes at each spatial-location. For such an ensemble, using a list decoder to decode a fraction of the RS component codes may partially achieve the threshold gains shown by our analysis in Sec. \ref{sec:genCompDec}. The added complexity and latency may be reduced by amortizing over the decoding window, e.g. only apply list decoding at the newest received block. List decoding may also incorporate soft information, either from the channel or between component-decoding iterations, to further improve performance.
%
%

\appendix

\subsection{Proof of Theorem \ref{thm:decoupInt}}\label{append:decoupInt}

We enumerate the number of decoupling interleavers. First, label the constraint half-edges entering the interleaver according to edge-type. For example, assign labels $\{1,\dots,Mn_c/w\}$ to constraint half-edges of type $0$ (i.e. the type $0$ bundle), $\{Mn_c/w+1,Mn_c/w+2,\dots,2Mn_c/w\}$ to constraint half-edges of type $1$, and in general, $\tau Mn_c/w + 1 + [Mn_c/w]$ to constraint half-edges of type $\tau$. Next, arbitrarily label the $N$ variable nodes using elements from the set $\{v,2v,\dots,Nv\}$. Finally, label the half-edges of each variable node by replacing each variable node label $i$ by a random permutation of the set $\{i-(v-1),\dots,i\}$. The interleaver makes connections between constraint and variable half-edges with the same label. 

Since we assume $M$ is arbitrarily large, we can assume that $v$ divides $Mn_c/w$ (i.e. the number of constraint half-edges in a bundle). Then, for an interleaver constructed by the above procedure, each constraint half-edge bundle is connected to an integer number of variable nodes. Moreover, all half-edges of each variable node are connected to the same constraint half-edge bundle. Therefore, an interleaver constructed according to the above procedure is a decoupling interleaver. Conversely, given an arbitrary decoupling interleaver, partition the variable half-edges according to edge-type. Since all variable half-edges incident on the same variable node have the same type, this also partitions the variable nodes. By arbitrarily assigning labels from the set $\{v,2v,\dots,Nv\}$ to the partitioned variable nodes, one can verify that every decoupling interleaver can be obtained by following the above construction.

There are $N!(v!)^N$ possible interleavers under the above construction, out of $(vN)!$ interleavers in total. Evaluating the probability using upper and lower bounds on the factorial and upper-bounding $v!$ by $v^v$ gives
\begin{align*}
\textrm{Pr}\{\textrm{decoupling interleaver}\} &= \frac{N!(v!)^N}{(vN)!} \\
&< \frac{(v^vN)^N}{(vN)^{vN}} \sqrt{\frac{2\pi N}{2\pi vN}}  \frac{\exp{\left(-N+\frac{1}{12N}\right)}}{\exp{\left(-vN+\frac{1}{12vN+1}\right)}} \\
&=\left(\frac{1}{N^{v-1}}\right)^N \sqrt{\frac{1}{v}} \exp{\left((v-1)N + \frac{v-1}{12vN+1}\right)} \\
&=O\left(N^{-(v-1)N}\right).
\end{align*}

%
%

\subsection{Proof of Lemma \ref{lemma:trend}}\label{subsec:trend}

Since all constraint nodes with degree type $\underline{i} \in S$ are un-recoverable, they are not affected by the initial removal of the single recoverable edge. Hence, we only consider the effects of the removal of the $v-1$ edges incident on the recovered variable node.

Let $\mathcal{E}^u_{\tau}$ denote the event that $u$ out of $v-1$ removed variable edges are of type $\tau$. Conditioned on $\mathcal{E}^u_{\tau}$, the expected change in the degree distribution at $k+\tau$ is the same as in peeling decoding where a degree-1 check node is removed along with a corresponding variable node of degree $u$ \cite[Sec. III-B]{luby:2001}. The conditional expected change is given by
\begin{multline*}
\mathbb{E}(\mathcal{R}^{k+\tau}_{(\underline{i})}(l+1) - \mathcal{R}^{k+\tau}_{(\underline{i})}(l) |\Theta_l,\mathcal{E}^u_{\tau}) \\
= [- i_{\tau}\mathcal{R}^{k+\tau}_{(\underline{i})}(l)+ (i_{\tau}+1)\mathcal{R}^{k+\tau}_{(\underline{i} + \underline{1}_{\tau})}(l)] \frac{u}{E_{\tau}}.
\end{multline*}
\noindent
If none of the removed edges are of type $\tau$, then
\[
\mathbb{E}(\mathcal{R}^{k+\tau}_{(\underline{i})}(l+1) - \mathcal{R}^{k+\tau}_{(\underline{i})}(l) |\Theta_l,\mathcal{E}^u_{\gamma \neq \tau}) = 0.
\]

Let $e_{\gamma}\in[w]$, $\gamma\in[v-1]$ denote the type of the $\gamma$th removed variable node edge. Define $\underline{e} = (e_0,\dots,e_{v-2})$ and
\[
h_{\tau}(\underline{e}) \triangleq \sum_{\gamma\in[v-1]}\mathbb{I}[e_{\gamma}=\tau].
\]
The probability of obtaining a sequence of edges of types given by $\underline{e}$ under uniform interleaving is
\[
\textrm{Pr}(\underline{e}) = \frac{ \prod_{\gamma} \prod_{i\in[f_{\gamma}(\underline{e})]} ( E_{\gamma} - i  )}{\prod_{i\in[v-1]}(F-i)}.  
\]
The probability of the event $\mathcal{E}^u_{\tau}$ is given by
\begin{align*}
\textrm{Pr}(\mathcal{E}^u_{\tau}) &= 
\sum_{\underline{e} : f_{\tau}(\underline{e}) = u}\frac{\prod_{i\in[u]} (E_{\tau}-i) \prod_{\gamma\neq\tau}\prod_{i\in[f_{\gamma}(\underline{e})]}(E_{\gamma}-i)}{\prod_{i\in[v-1]}(F-i)} \\
&= \binom{v-1}{u}\frac{\prod_{i\in[u]}(E_{\tau}-i)\prod_{j\in[v-1-u]}(F-E_{\tau}-j)}{\prod_{i\in[v-1]}(F-i)}
\end{align*}
\noindent
The unconditional trend function is then
\begin{multline*}
\mathbb{E}(\mathcal{R}^{k+\tau}_{(\underline{i})}(l+1) - \mathcal{R}^{k+\tau}_{(\underline{i})}(l) |\Theta_l) = \\
\sum_{u=1}^{v-1} [- i_{\tau}\mathcal{R}^{k+\tau}_{(\underline{i})}(l)+ (i_{\tau}+1)\mathcal{R}^{k+\tau}_{(\underline{i} + \underline{1}_{\tau})}(l)] \frac{u}{E_{\tau}} \textrm{Pr}(\mathcal{E}^u_{\tau}) \\
= [- i_{\tau}\mathcal{R}^{k+\tau}_{(\underline{i})}(l)+ (i_{\tau}+1)\mathcal{R}^{k+\tau}_{(\underline{i} + \underline{1}_{\tau})}(l)]\frac{1}{E_{\tau}}\\
\times \frac{(v-1)E_{\tau}\prod_{i=1}^{v-2}(F-i)}{\prod_{i\in[v-1]}(F-i)} \\
= [- i_{\tau}\mathcal{R}^{k+\tau}_{(\underline{i})}(l)+ (i_{\tau}+1)\mathcal{R}^{k+\tau}_{(\underline{i} + \underline{1}_{\tau})}(l)]\frac{v-1}{F}
\end{multline*}
where the sum over $u$ is tedious but straightforward to evaluate using induction on $v$.

\subsection{Potential function analysis}\label{append:potFcn}

\begin{definition}\label{def:admissible}
\cite[Definition 27]{pfister:2014} An admissible system is a coupled scalar recursion where the functions $f(x,\lambda),g(x)$ satisfy 
\begin{enumerate}
\item $f(x,\lambda)$ is a $C^1$ function on $\mathcal{X}\times\Lambda$ and $g(x)$ is a $C^1$ function on $\mathcal{X}$
\item $f(x,\lambda)$ is non-decreasing in $x$ and $\lambda$ 
\item $g(x)$ is strictly increasing in x
\item $\frac{d^2}{dx^2}g(x)$ exists and is continuous on $\mathcal{X}$.
\end{enumerate}
Let $h(x,\lambda)=f(g(x),\lambda)$ denote the single-system (uncoupled) recursion. An admissible system is proper if $\frac{\partial}{\partial\lambda}h(x,\lambda)>0$ for all $(x,\lambda)\in \mathcal{X}\setminus 0 \times \Lambda$.
\end{definition}


\begin{definition}[Single-system potential]\label{defn:pot_fcn} 
\cite[Eqn. 15]{pfister:2014}
The single-system or uncoupled potential function of an admissible system $f(x,\lambda)$, $g(x)$ is defined as 
\begin{equation*}
U_s(x,\lambda) = xg(x)-\int_0^xg(z)\,dz-\int_0^{g(x)}f(z,\lambda)\,dz.
\end{equation*}
\end{definition}

\begin{definition}[Potential threshold]\label{defn:pot_thres}
\cite[Definitions 28, 30, 31]{pfister:2014} Let
\begin{align*}
\Psi(\lambda) &= \min_{x\in\mathcal{X}}U_s(x,\lambda) \\
X^*(\lambda) &= \{x\in\mathcal{X}|U_s(x,\lambda)=\Psi(\lambda)\} \\
\bar{x}^*(\lambda) &= \max X^*(\lambda). 
\end{align*}

The potential threshold $\lambda^*$ is defined as
\begin{equation}\label{eqn:lambda_star}
\lambda^*=\sup\{\lambda\in\Lambda|\bar{x}^*(\lambda)=0\}
\end{equation}

\end{definition}

The stability threshold $\lambda_{\textnormal{stab}}^*$ is defined as
\begin{equation*}
\lambda^*_{\textnormal{stab}}=\sup\{\lambda\in\Lambda|\exists\delta>0,\forall x\in(0,\delta],h(x,\lambda)<x\}.
\end{equation*}
\noindent
An admissible system is called unconditionally stable if there is a $\delta>0$ such that $h(x,\lambda_{\textrm{max}})<x$ for all $x\in(0,\delta]$.
\begin{lemma}\label{lemma:pot_thres}
\cite[Theorem 1, Lemma 32(iv)]{pfister:2014} If $\lambda<\lambda^*\leq\lambda^*_{\textnormal{stab}}$ then there is a $w_0<\infty$ such that the admissible system converges to $\mathbf{0}$ for all $w>w_0$.
\end{lemma}
\noindent
For an unconditionally stable coupled scalar recursion, the potential threshold is only determined by $\lambda^*$. By the above lemma, the potential threshold is the ensemble threshold with $W\rightarrow\infty$ and $w>w_0$. Recall that we assume $M\rightarrow\infty$ in the analysis. The potential threshold is thus an upper-bound on the threshold of the spatially-coupled split-component ensemble. The potential threshold can also be defined via the \emph{fixed-point potential}.

\begin{definition}[Fixed point potential]\label{defn:fp_pot}
\cite[Definition 37]{pfister:2014}
For admissible systems, define the fixed-point support set
\begin{equation*}
\mathcal{X}_f = \{x\in\mathcal{X}\setminus 0\, |\, h(x;0)\leq x,\, h(x,\lambda_{max})\geq x\}.
\end{equation*}
Let $\tilde\lambda(x) = \min{\{\lambda\in\Lambda\, |\, h(x,\lambda)=x,\,x\in\mathcal{X}_f\}}$, the fixed-point potential is a function $Q:\mathcal{X}_f\mapsto\mathbb{R}$
\begin{equation*}
Q(x)=U_s(x,\tilde\lambda(x)).
\end{equation*}
\end{definition}

\begin{lemma}\label{lemma:fp_pot}
\cite[Lemma 41(iii), 42(iii)]{pfister:2014}
For an unconditionally stable admissible system, if $\lambda^*\in[0,\lambda_{\textnormal{max}})$ then
\begin{equation*}
\lambda^* = \min \{ \tilde\lambda(x)\, |\, x\in\overline{\mathcal{X}_f},\, Q(x)=0 \}
\end{equation*}
\noindent
where $\overline{\mathcal{X}_f}$ is the closure of $\mathcal{X}_f$.
\end{lemma}

\bibliographystyle{IEEEtran}
\bibliography{IEEEabrv,scsc_it_draft.bib}

\end{document}